\documentclass[useAMS,usenatbib]{mn2e}
\usepackage{epsfig}
\usepackage{multicol}

\title[Protostellar discs formed from turbulent cores]
{Protostellar discs formed from turbulent cores}
\author[S.~Walch, T.~Naab, A.~Burkert, A.~Whitworth, M.~Gritschneder]{S. Walch$^{1,2}$\thanks{E-mail: Stefanie.Walch@astro.cf.ac.uk}, T.~Naab$^{1}$, A.~Burkert$^{1}$, A.~Whitworth$^{2}$, M.~Gritschneder$^{1}$\\
$^{1}$University Observatory, University of Munich, Scheinerstr. 1, Munich, 81679, Germany\\
$^2$School of Physics \& Astronomy, Cardiff University, 5 The Parade, Cardiff CF24 3AA, Wales, UK}
\begin{document}

\date{Accepted . Received 2009 January 13; in original form }

\pagerange{\pageref{firstpage}--\pageref{lastpage}} \pubyear{2009}

\maketitle

\label{firstpage}

\begin{abstract}
We investigate the collapse and fragmentation of low-mass, trans-sonically turbulent prestellar cores, using SPH simulations. The initial conditions are slightly supercritical Bonnor-Ebert spheres, all with the same density profile, the same mass ($M_{_{\rm O}}=6.1\,{\rm M}_\odot$) and the same radius ($R_{_{\rm O}}=17,000\,{\rm AU}$), but having different initial turbulent velocity fields. Four hundred turbulent velocity fields have been generated, all scaled so that the mean Mach number is ${\cal M}=1$. Then a subset of these, having a range of net angular momenta, $j$, has been evolved. The evolution of these turbulent cores is not strongly correlated with $j$. Instead it is moderated by the formation of filamentary structures due to converging turbulent flows. A high fraction ($\sim 82\%$) of the protostars forming from turbulent cores are attended by protostellar accretion discs, but only a very small fraction ($\sim 16\%$) of these discs is sufficiently cool and extended to develop non-linear gravitational instabilities and fragment.
\end{abstract}


\section{Introduction}

One of the key unsolved problems in star formation is how a low-mass prestellar core is converted into a small stellar system, and what role is played in this process by protostellar accretion discs. This problem is crucial to understanding how the core mass function \citep[e.g.][]{Motte1998, Testi1998, Johnstone2001, Nutter2007, Alves2007} maps into the system and stellar initial mass functions \citep{Kroupa2001a, Chabrier2003}.
In a companion paper \citep[][ hereafter W09]{Walch2009} we have simulated the evolution of {\it rigidly rotating} cores having different net angular momenta. Such cores collapse to produce a primary protostar surrounded by a protostellar disc, but only cores with high angular momenta produce protostellar discs which are sufficiently cool and extended to fragment and thereby spawn secondary companions. In this paper we simulate the collapse and fragmentation of {\it mildly turbulent} prestellar cores having the same range of net angular momenta as those simulated in W09. These mildly turbulent cores tend to fragment in a completely different way: first a filament forms, and then protostars precipitate out of the filament.

The context for this study is the paradigm in which molecular clouds are assembled, form stars and disperse, on a dynamical timescale \citep[e.g.][]{Elmegreen2000, Vazquez2000, Burkert2004}. In this paradigm the turbulent energy delivered by the assembly of the molecular cloud is dissipated as it cascades to smaller scales\footnote{Additional turbulent energy may be injected by radiative and mechanical feedback from stars, but Larson's scaling relations (as discussed in Section 2.2) strongly suggest that most turbulent energy is injected on the largest scales and cascades to smaller scales}. Thus the larger, more diffuse structures within molecular clouds, those that form star clusters, have supersonic turbulence. This is the regime that has been simulated by \citet{Bonnell2008} and \citet{Bate2009}. Conversely, the smaller, denser prestellar cores, which form single stars and small stellar systems, display trans-sonic or subsonic turbulence \citep[e.g. ][]{Myers1991, Andre1993, Barranco1998}. This is the regime with which the present paper is concerned.

Previous numerical work on the collapse and fragmentation of turbulent low-mass cores has concentrated on the statistical properties of the resulting protostars (i.e. masses and multiplicities) rather than the phenomenology of fragmentation. For example, \citeauthor{Goodwin2004} (\citeyear{Goodwin2004a, Goodwin2004b, Goodwin2006}) follow the collapse and fragmentation of $5.4\,{\rm M}_\odot$ cores having Plummer-like density profiles and very low initial levels of turbulence characterised by $\gamma_{\rm TURB}=0.05,\,0.10,\,{\rm and}\,0.25$, where
\begin{eqnarray}
\gamma_{_{\rm TURB}}&=&\frac{U_{_{\rm TURB}}}{|U_{_{\rm GRAV}}|}\,;
\end{eqnarray}
here $U_{_{\rm TURB}}$ and $U_{_{\rm GRAV}}$ are the initial turbulent and gravitational energies. Goodwin et al. use SPH with sink particles and a barotropic equation of state, and perform multiple realisations for each value of $\gamma_{_{\rm TURB}}$. On average these simulations produce one star per initial Jeans mass, and $80\%$ of cores spawn multiple systems. Recently, \citet{Attwood2009} have repeated these simulations, solving the energy equation explicitly, and using an approximate method to treat the associated transfer of cooling radiation \citep[see ][]{Stamatellos2007}. These simulations tend to produce more protostars -- and in particular more very low-mass protostars -- than the earlier simulations performed using a barotropic equation of state.

\citet{Offner2008} model the collapse of turbulent cores, using AMR and a barotropic equation of state. Their cores are produced in a large-scale simulation as part of a molecular cloud, with either driven or decaying turbulence. Individual cores with masses and sizes comparable to those treated here are then followed at higher resolution; this has the advantage that the cores have consistent velocity and density fields from the outset. Their principal finding is that simulations with decaying turbulence form more low-mass protostars than simulations with driven turbulence, because the decline in turbulent support allows forming protostars to fall into the centre of the core where they then experience competitive accretion and dynamical ejections.

The plan of this paper is as follows. In Section 2 we describe the initial conditions, the numerical code and the constitutive physics; we concentrate on the motivation for, and the generation of, the initial velocity field, since this is the only aspect that is different from W09; more detail on the other aspects can be found in W09. In Sections 3 through 5, we present the results, and discuss the role played by filaments and discs in the formation and growth of protostars. In Section 6 we summarise our main conclusions.

\section{Initial conditions, numerical method, and constitutive physics}

\subsection{The initial core density profile}\label{SEC:BES}

As in W09, we model cores as Bonnor-Ebert spheres truncated at dimensionless radius $\xi_{_{\rm B}}=6.9$, and then we increase the density everywhere by $10\%$. The gas is pure H$_{_2}$ at $28\,{\rm K}$, so the isothermal sound speed is $a_{_{\rm O}}=0.34\,{\rm km}\,{\rm s}^{-1}$. The central density is $\rho_{_{\rm C}}=10^{-18}\,{\rm g}\,{\rm cm}^{-3}$, so a core has mass $M_{_{\rm O}}\simeq 6.1{\rm M}_\odot$, initial radius $R_{_{\rm O}}\simeq 17,000\,{\rm AU}$, boundary pressure $P_{_{\rm EXT}}\simeq k_{_{\rm B}}\,5.5\times 10^5\,{\rm cm}^{-3}\,{\rm K}$, and ratio of thermal to gravitational energy $\alpha=0.74$. The freefall time at the centre is $67\,{\rm kyr}$ and the sound crossing time is $250\,{\rm kyr}$.

\subsection{The initial velocity field}\label{SEC:VEL}

The velocity fields in molecular clouds appear to subscribe (with some scatter) to a scaling law of the form $\sigma(\lambda)\propto\lambda^q$, where $\lambda$ is the linear size of a coherent feature, and $\sigma(\lambda)$ is its internal velocity dispersion. If these velocities derive from turbulence with a power law $P_k\propto k^n$, then $n=-3-2q$ \citep{Myers1999}. \citet{Larson1981} finds $q\simeq 0.38$, corresponding to $n=-3.76$ (close to the $n=11/3=3.67$ of Kolmogorov 1936). \citet{Goodman1998} find $q\simeq 0.5$ on scales $\lambda\ga 0.1\,{\rm pc}$, corresponding to $n=-4$.

Cores created in a turbulent molecular cloud will have irregular internal velocity fields, and in general they will inherit a net angular momentum \citep{Goldsmith1985, Dubinski1995, Goodman1993, Barranco1998, Caselli2002}. \citet{Jijina1999} estimate that a low-mass core typically has a ratio of turbulent to gravitational energy in the range $0<\gamma_{_{\rm TURB}}\la 0.5$, and mean specific angular momentum $\bar{j}\sim 10^{21}\,{\rm cm}^2\,{\rm s}^{-1}$. \citet{BurkertBodenheimer00} have shown that these features can be reproduced if the turbulence has $n=-3$ or $n=-4$.

We have therefore created turbulent velocity fields with the same ansatz as \citet{BurkertBodenheimer00}. First we generate a random Gaussian velocity field with $P_k\propto k^{-4}$, populating the wavenumbers $k_{_{\rm MIN}}\leq k\leq k_{_{\rm MAX}}$, with $k_{_{\rm MIN}}=1$ (corresponding to the diameter of the cloud) or $k_{_{\rm MIN}}=2$ (corresponding to the radius of the cloud), and $4\leq k_{_{\rm MAX}}\leq 8$. Then we map this field onto a uniform $128^3$ grid, covering the whole computational domain. Next we scale the velocities of the grid points so that their mean Mach number is unity (i.e. the turbulence is transonic). Finally, we obtain the velocities of individual SPH particles by interpolating on this grid.

\begin{table}\label{TAB:SIMUS}
\begin{center}
\begin{tabular}{cccccccc}\hline
$\!\!\!{\rm Run}\!\!\!$ & $j$ & $\gamma_{_{\rm TURB}}$ & $t_{_{\rm O}}$ & $t_{_{\rm END}}$ & $M_\star$ & $M_{_{\rm D}}$ & $R_{_{\rm D}}$\\
 & $\!\!\overline{10^{21}{\rm cm}^2\!/{\rm s}}\!\!$ & & $\overline{\rm kyr}$ & $\overline{\rm kyr}$ & $\overline{{\rm M}_\odot}$ & $\overline{{\rm M}_\odot}$ & $\overline{AU}$\\\hline
1b & 1.03 & 0.410 & 151 & 163 & 0.29 & 0.72 & 327\\
1d & 1.03 & 0.284 & 110 & 116 & 0.37 & 0.24 & 310\\
1f & 1.03 & 0.393 & 135 & 148 & 0.38 & 0.42 & 213\\
2b & 1.37 & 0.466 & 137 & 141 & 0.40 & 0.70 & 1572\\
3a & 1.71 & 0.327 & 126 & 136 & 0.33 & 0.62 & 274\\
4b & 2.59 & 0.630 & 159 & 180 & 0.26 & 0.40 & 250\\
6d & 2.72 & 0.483 & 112 & 138 & 0.22 & 0.85 & 676\\\hline  
\end{tabular}
\end{center}
\caption{Column 1 gives the ID of the simulation; the number corresponds to the ID of the rigidly rotating simulation having the same net angular momentum, and the lower-case letter identifies which realisation of the velocity field has been adopted. Column 2 gives the mean specific angular momentum, $j$. Column 3 gives the ratio of rotational to gravitational energy, $\gamma_{\rm TURB}$. Columns 4 and 5 give the times at which the primary protostar forms and the simulation is terminated. Columns 6 and 7 give the final masses of the primary protostar and attendant disc. Column 7 gives the radial extent of the disc.}
\end{table}

Four hundred velocity fields have been generated in this way. We have then selected a subset of these velocity fields having the same -- or very nearly the same -- net angular momenta as the rotating cores simulated in W09. Key parameters of this subset are presented in Table 1; these are the cores whose evolution we have simulated, and whose evolution we discuss in this paper. We find that only when we use $k_{_{\rm MIN}}=1$ (i.e. there is turbulent energy on the scale of the diameter of the core), does the resulting distribution of core angular momenta match the observations, with $\bar{j}\sim 10^{21}\,{\rm cm}^2\,{\rm s}^{-1}$. If we set $k_{_{\rm MIN}}=2$ (i.e. the largest turbulent wavelength is the core radius), then the mean specific angular momentum is an order of magnitude lower, $\bar{j}\sim 10^{20}\,{\rm cm}\,{\rm s}^{-1}$. 

\subsection{The VINE code}\label{SEC:CODE}

As in W09, the simulations have been performed with the Tree-SPH code VINE. VINE is described fully in \citet{Wetzstein2008} and \citet{Nelson2008}. It is parallelised with OpenMP directives. It invokes a leapfrog integrator, and individual particle time steps. We adopt a CFL tolerance parameter of 0.1. Gravitational accelerations are estimated using a tree, with opening angle $\theta=0.005$ \citep{Springel2001}. The gravitational softening length is set to the hydrodynamical smoothing length \citep{BateBurkert97}, which is adapted so that each particle has ${\cal N}_{_{\rm NEIB}}=50\pm20$ neighbours. Hydrodynamical forces are treated with periodic boundary conditions, but gravitational forces are not. Artificial viscosity is treated using the standard prescription of \citet{Gingold1983} with $\alpha_{_{\rm AV}}=1$ and $\beta_{_{\rm AV}}=2$, plus the Balsara switch \citep{Balsara1995} to ensure a better treatment of shear flows.

\subsection{Equation of state and molecular line cooling}\label{SEC:ENEQN}

As in W09, we compute the gross thermodynamics on the assumption that the gas is pure H$_{_2}$, with ratio of specific heats $\gamma=7/5$ and molecular weight $\mu=2$. We solve the energy equation, with the following prescription for the radiative cooling. At low densities, $\rho<\rho_{_{\rm CRIT}}=10^{-13}\,{\rm g}\,{\rm cm}^{-3}$, we use the cooling rates computed by \citet{Neufeld1995}, with the constraint that the temperature is not allowed to fall below $T=9\,{\rm K}$. At high densities, $\rho>\rho_{_{\rm CRIT}}$, we switch off the radiative cooling, and the gas then evolves adiabatically (but not isentropically). The same procedure was adopted by \citet{Banerjee2004}.

\subsection{Resolution}\label{SEC:RES}

As in W09, the core is modeled with 430,000 SPH particles, each having mass $m_{_{\rm SPH}}=1.4\times 10^{-5}\,{\rm M}_\odot$. Hence the minimum resolvable mass is $2{\cal N}_{_{\rm NEIB}}m_{_{\rm SPH}}=1.4\times 10^{-3}\,{\rm M}_\odot$ and the Jeans mass is always resolved. In addition there are 24,000 ambient particles exerting the external pressure that contains the core. We do not use sink particles. Instead we impose a minimum smoothing length, $h_{_{\rm MIN}}=2\,{\rm AU}$.

\subsection{Definitions}

As in W09, the protostars in these simulations comprise all material having density $\rho>10^{-11}\,{\rm g}\,{\rm cm}^{-3}$, and the first protostar to form is the primary protostar. Similarly the attendant protostellar discs comprise all material having density in the range $10^{-16}\,{\rm g}\,{\rm cm}^{-3}<\rho<10^{-11}\,{\rm g}\,{\rm cm}^{-3}$; \citet{Offner2009} use a similar definition for discs. Filaments are then defined as all material with density $\rho<10^{-16}\,{\rm g}\,{\rm cm}^{-3}$ and temperature $T>20\,{\rm K}$. The material in a filament is warm, because it is heated in an accretion shock as it enters the filament. We use the variables $r=\left(x^2+y^2+z^2\right)^{1/2}$ and $\omega=\left(x^2+y^2\right)^{1/2}$ to distinguish, respectively, distance from the centre of co-ordinates and distance from the $z$-axis. When analysing protostellar discs, we shift the co-ordinate system so that the protostar is at $(x,y,z)=(0,0,0)$ and $z$ points along the disc's shortest axis of inertia. 

\section{Overview of simulations}
\begin{figure*}
  \begin{multicols}{2}
  \epsfig{figure=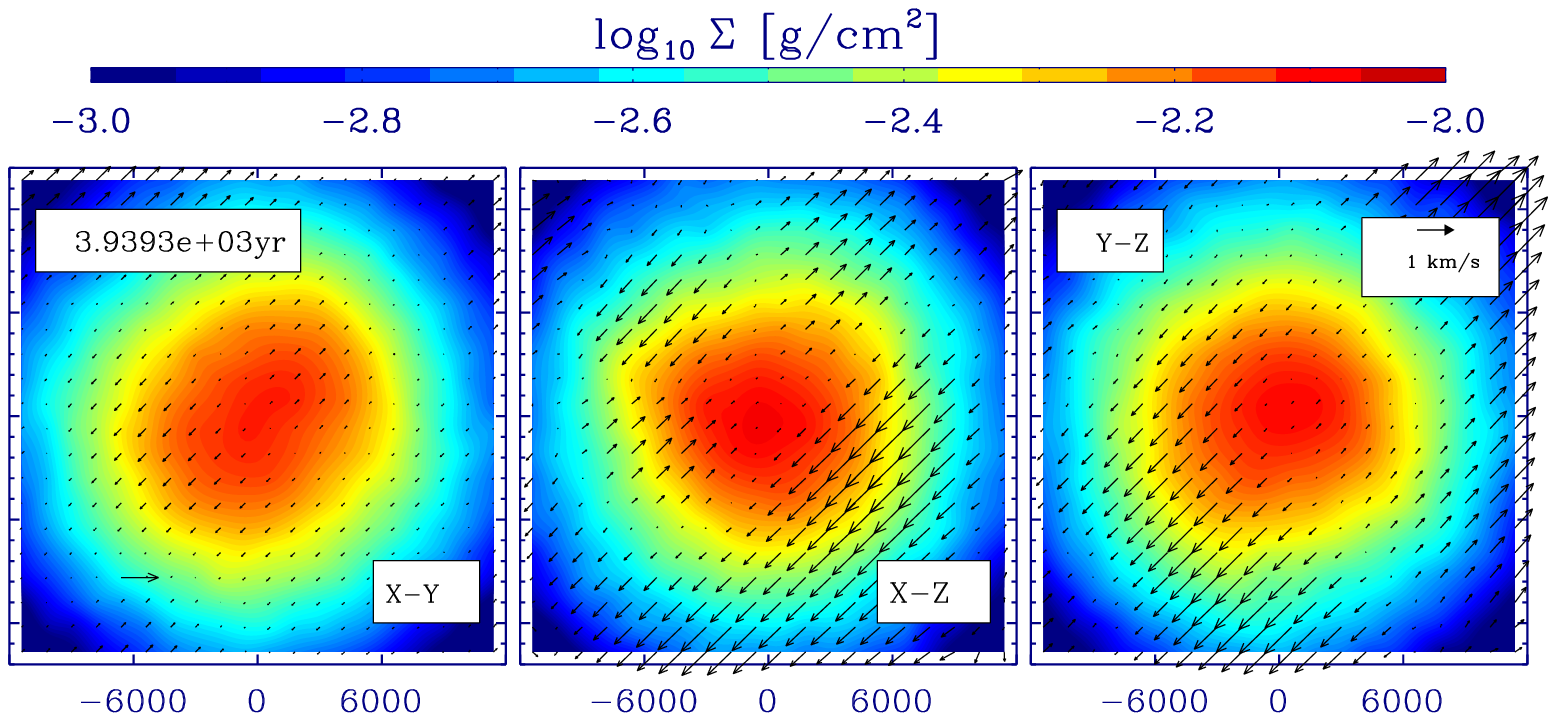 ,width=9cm}
  \epsfig{figure=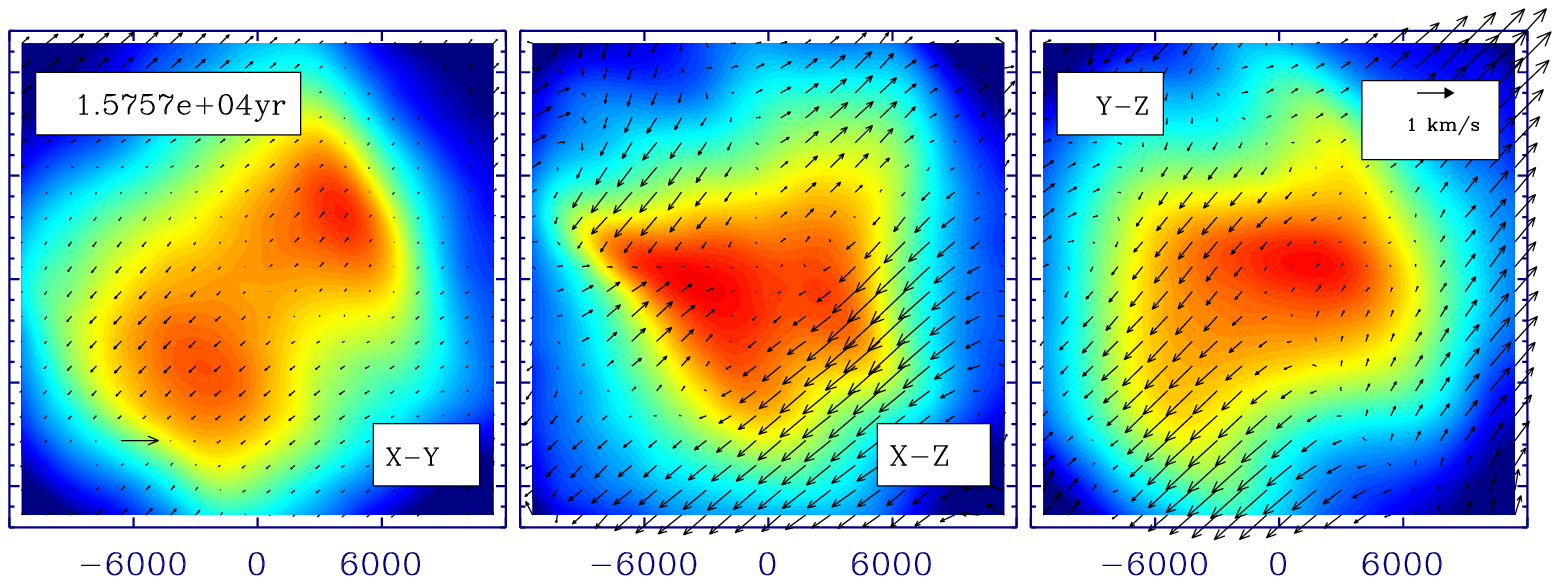 ,width=9cm}
  \epsfig{figure=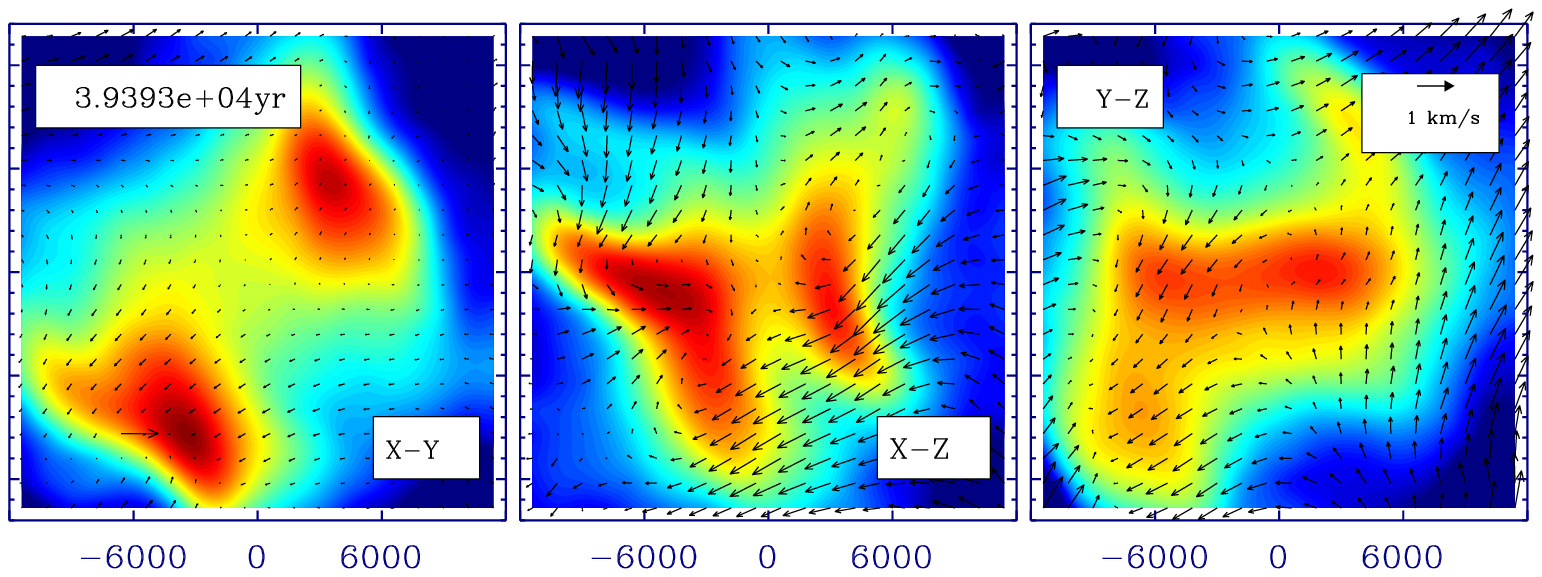 ,width=9cm}
  \epsfig{figure=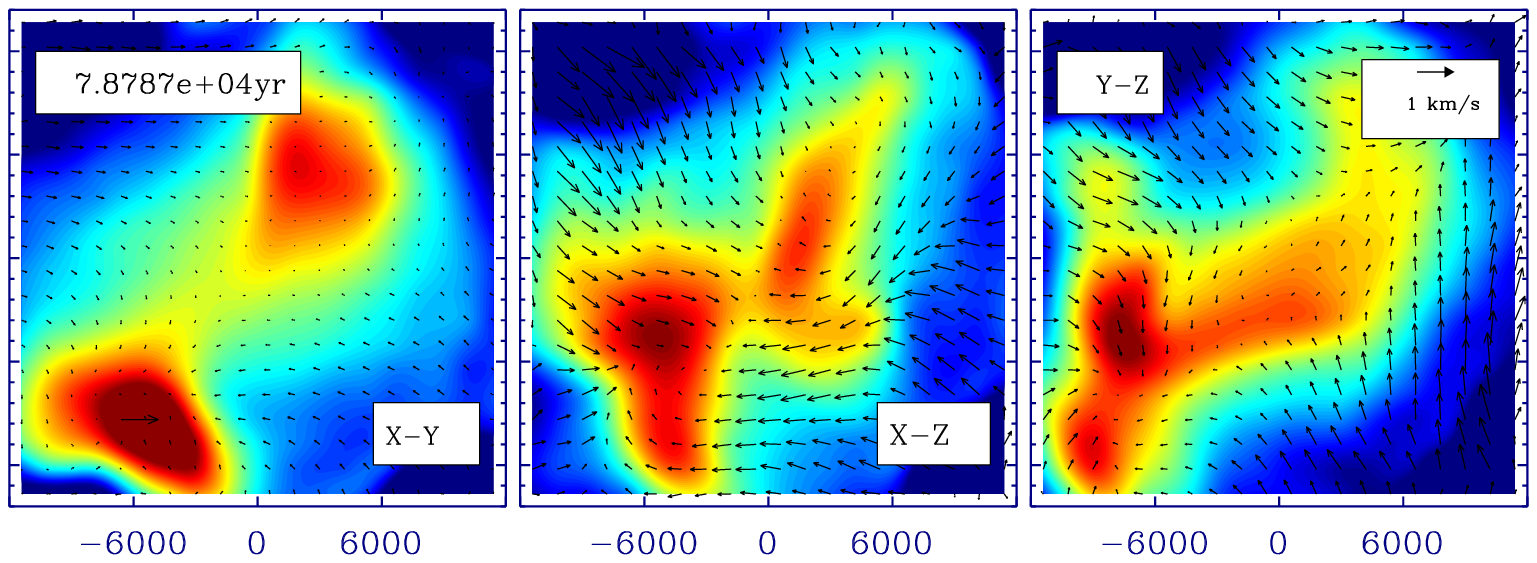 ,width=9cm}
  \epsfig{figure=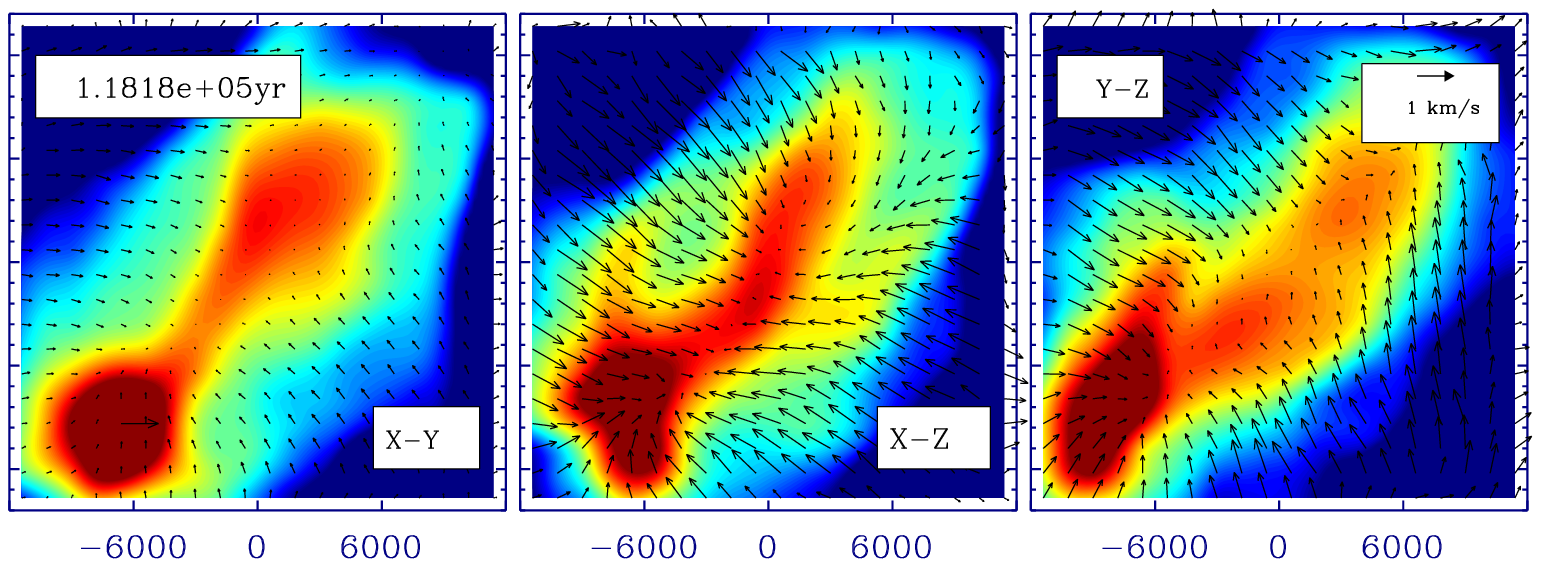 ,width=9cm}
  \epsfig{figure=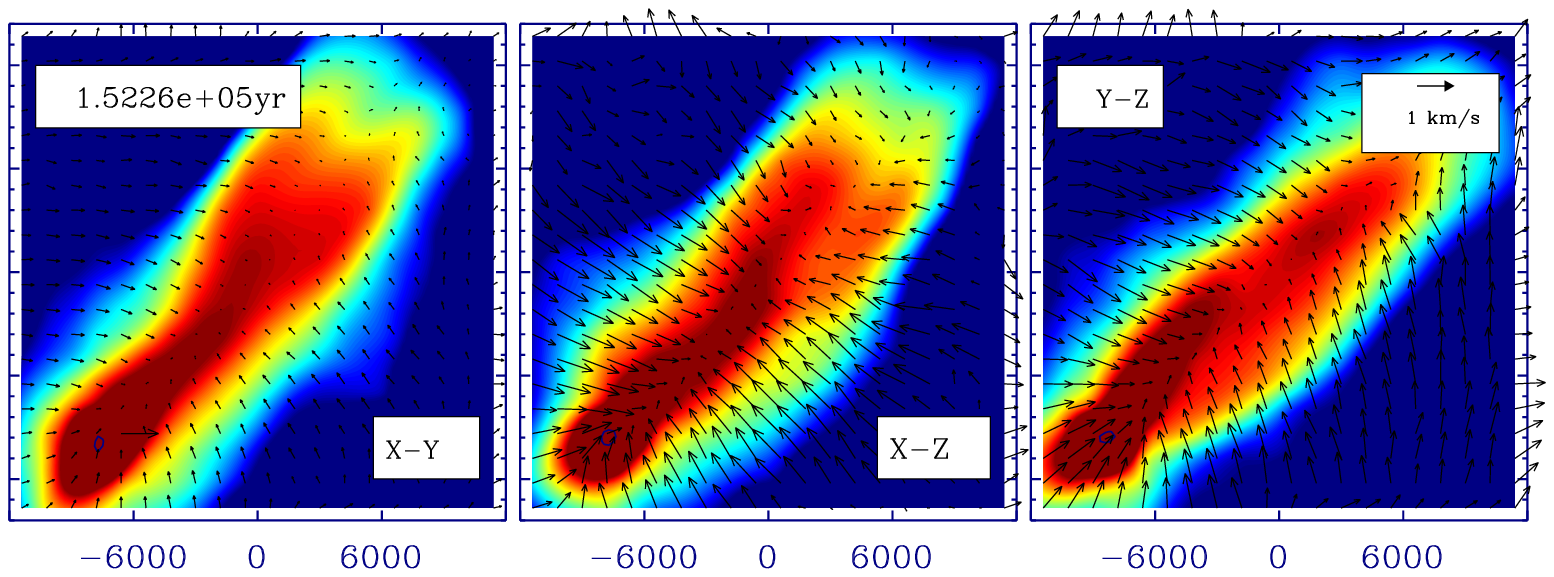 ,width=9cm}\\
  \epsfig{figure=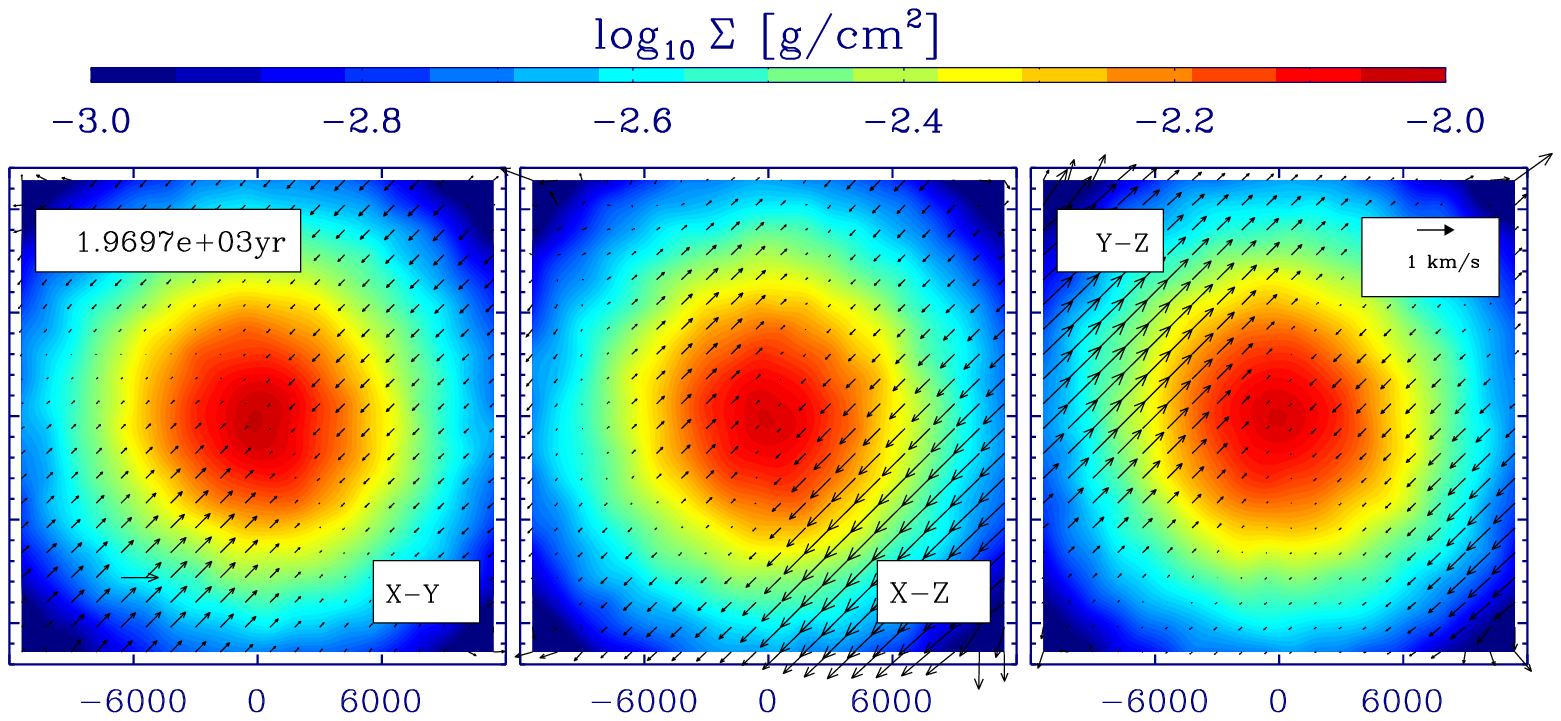 ,width=9cm}
  \epsfig{figure=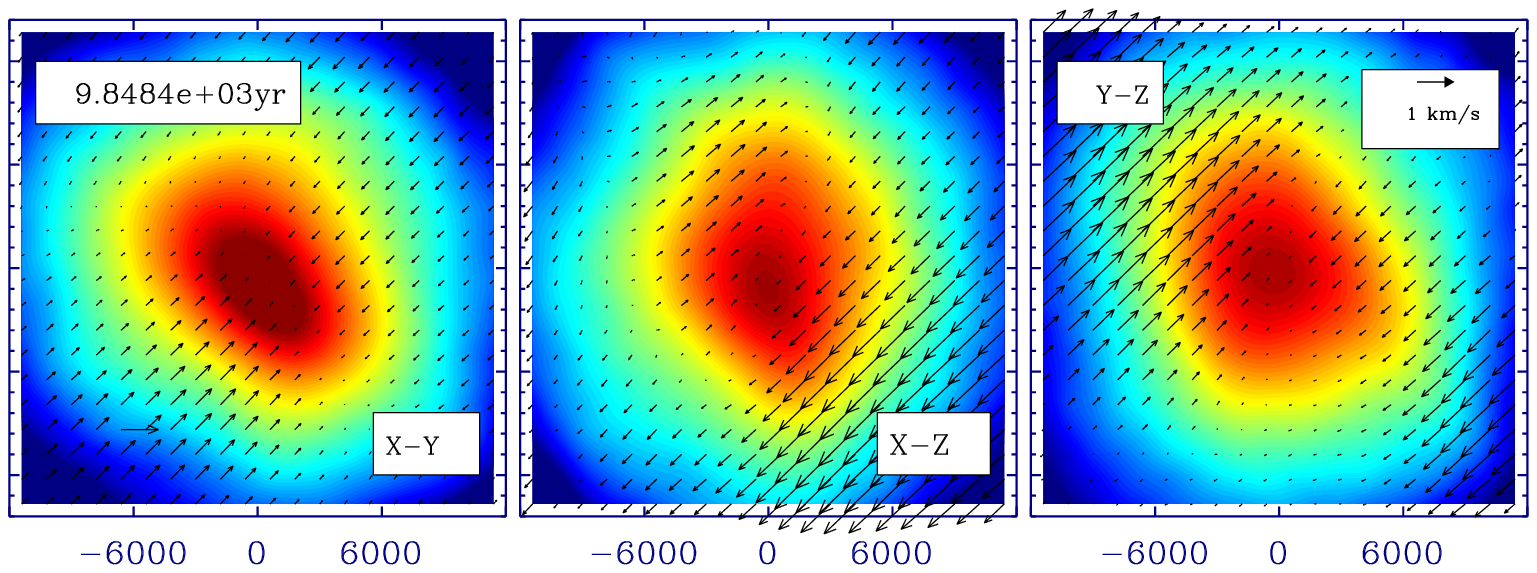 ,width=9cm}
  \epsfig{figure=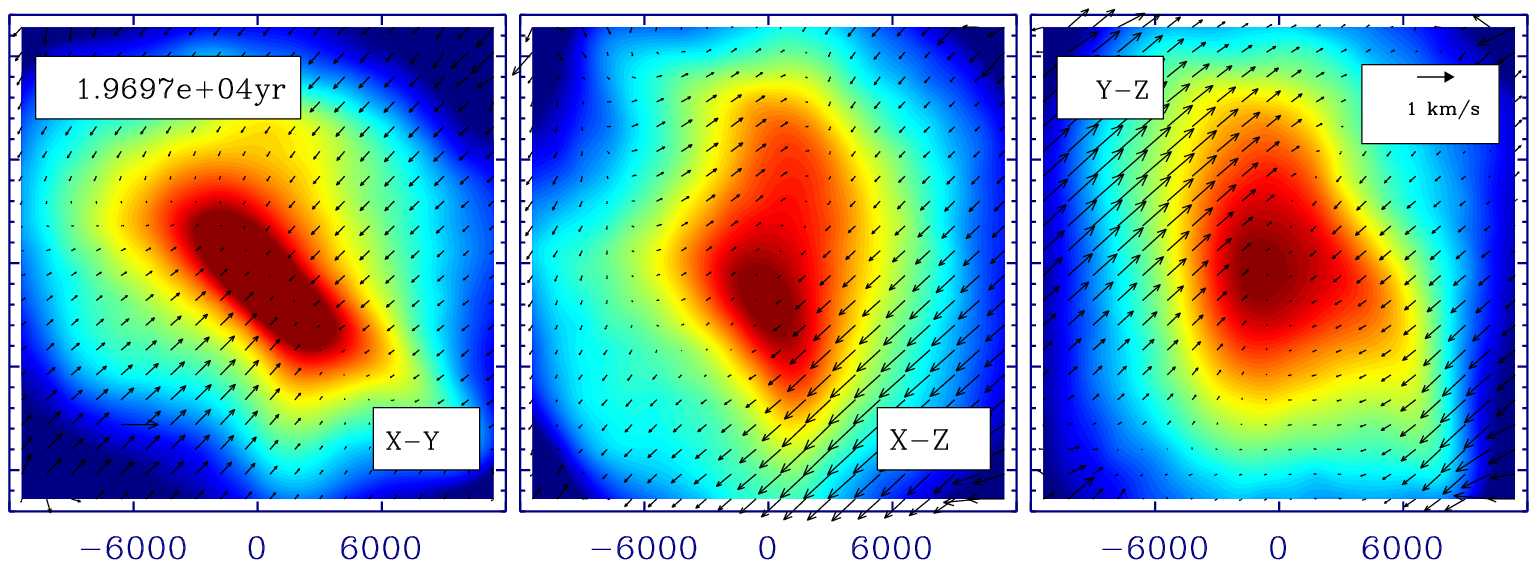 ,width=9cm}
  \epsfig{figure=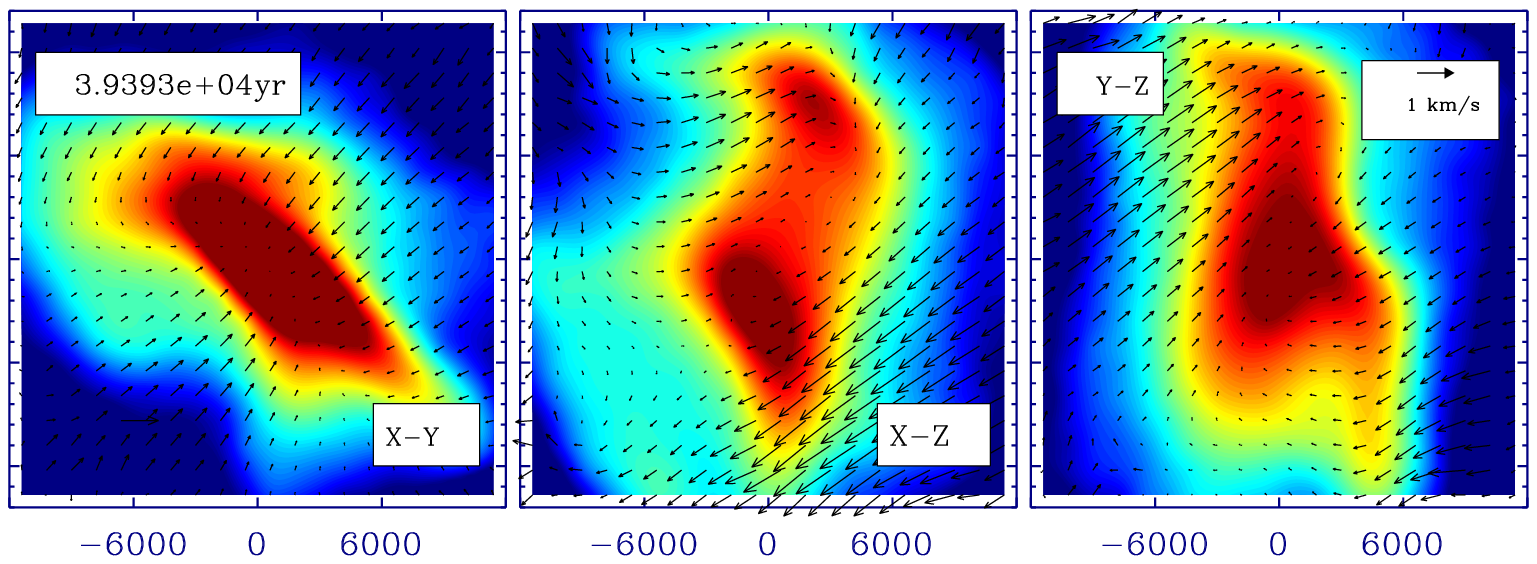 ,width=9cm}
  \epsfig{figure=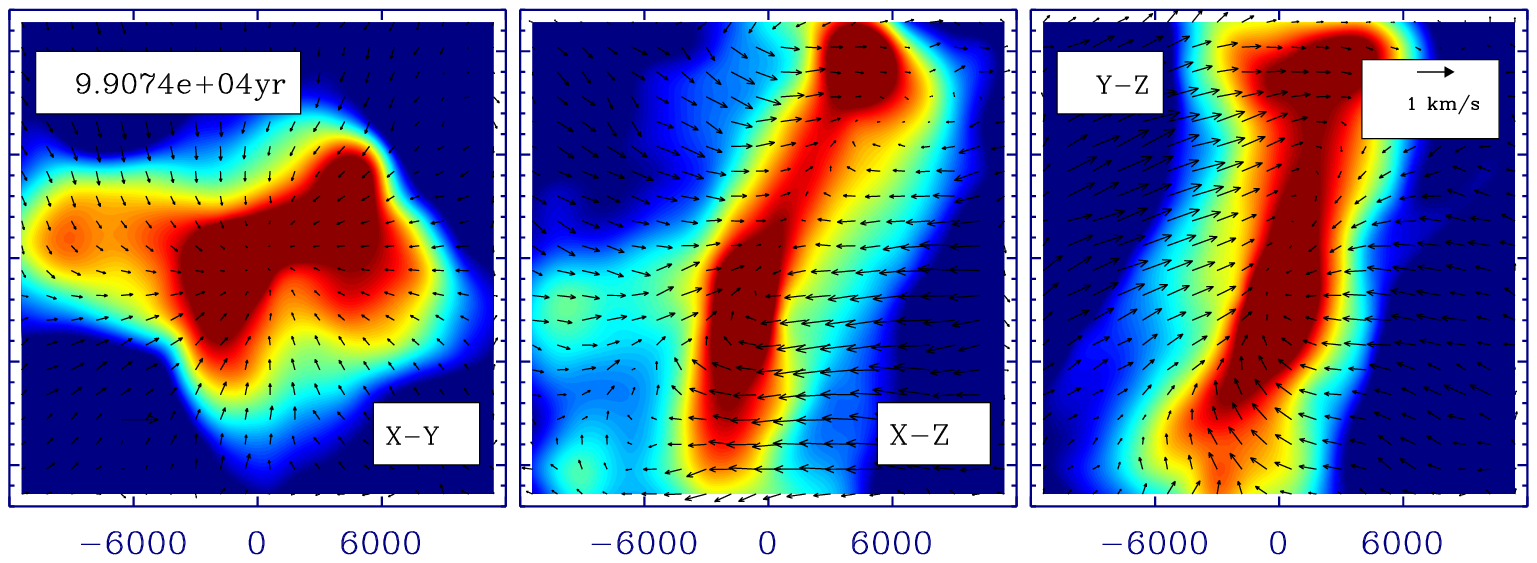 ,width=9cm}
  \epsfig{figure=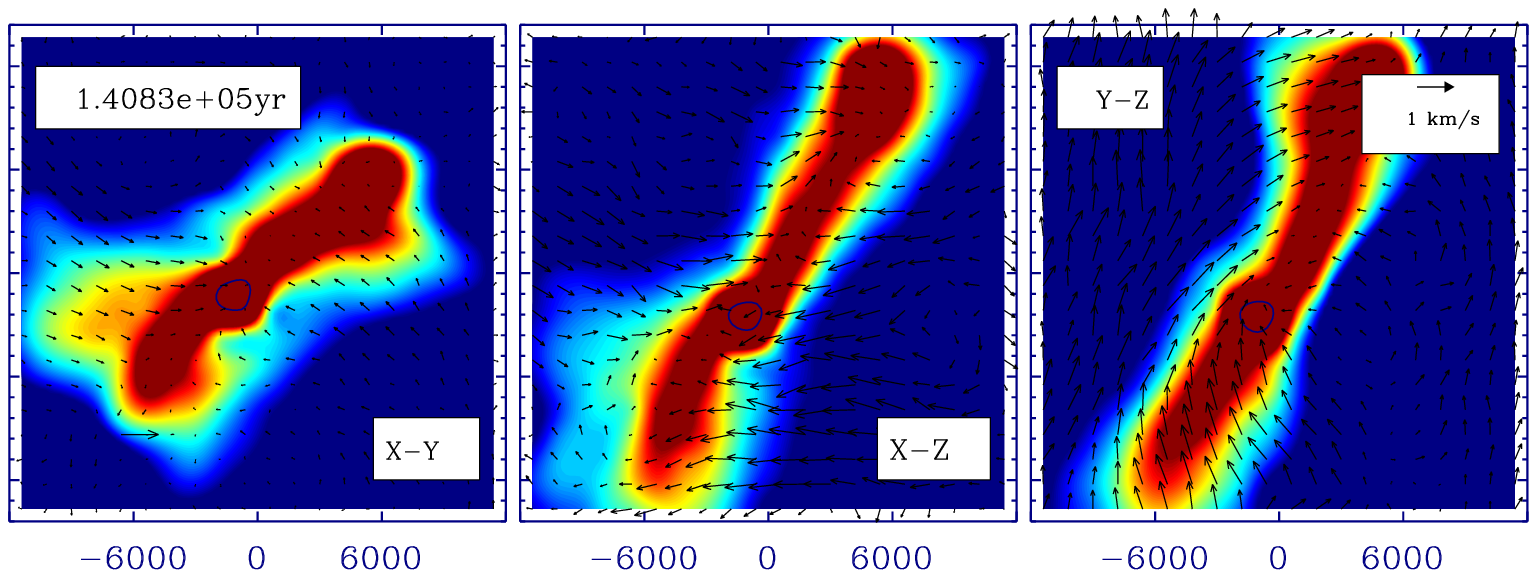 ,width=9cm}
  \end{multicols}
\caption{False-colour column-density images for Runs 1b and 6d, with column-averaged velocity vectors superimposed. The three lefthand columns show frames from Run 1b; successive rows correspond to the times $t=3.94,\,15.76,\,39.39,\,78.79,\,118.2\;{\rm and}\;152.3\;{\rm kyr}$. The three righthand columns show frames from Run 6d; successive rows correspond to the times $t=1.97,\,9.85,\,19.70,\,39.39,\,99.07\;{\rm and}\;140.8\;{\rm kyr}$. The density scale is shown at the top of the page, and a $1\,{\rm km}\,{\rm s}^{-1}$ velocity vector is given on the $x=0$ slices. All frames are $(24000\,{\rm AU})^2$.}\label{FIG:FILAMENTS}
\end{figure*}

{\sc the effect of the turbulent velocity field.} In turbulent simulations, regions of large $|\nabla\cdot{\bf v}|$ are present throughout the volume of the core, from the outset, and the spherical symmetry of the core is therefore immediately broken. This means that at the start of the simulation, some parts of the core are expanding ($\nabla\cdot{\bf v}>0$) and other parts are contracting ($\nabla\cdot{\bf v}<0$). Some of the expanding parts may even expand forever and disperse, but most remain bound and eventually fall back towards the centre of mass. The contracting parts tend to form coherent filamentary structures, and once these filaments become self-gravitating, protostellar fragments condense out of them. Sometimes two neighbouring condensations form from the same filament and subsequently merge to form a single protostar. On other occasions two condensations form at well separated locations, and in this case they are likely to condense separately and form a wide binary protostar. However, such wide binary protostars appear to be quite rare.

{\sc the location of protostar formation.} Protostellar condensations are not in general centrally located within their birth core, and in one simulation a protostar actually forms outside the initial boundary of the birth core (Run 4b). Typically a protostar has a peculiar velocity of $\sim 0.1\,{\rm km}\,{\rm s}^{-1}$, relative to the birth core. This is consistent with the results of \citet{Beichman1986}, who looked at the locations of IRAS sources in the vicinity of dense NH$_3$ cores having comparable masses and radii to those simulated here. About 40\% of the IRAS sources had visible counterparts (e.g. T Tauri stars) that lay just outside their associated core, implying either that they formed there, or -- more probably -- that they migrated there. During a mean lifetime of $10^6\,{\rm yr}$ and given a mean peculiar velocity of $0.1\,{\rm km}\,{\rm s}^{-1}$, most mature T Tauri stars would be found just outside their birth cores.

{\sc protostellar discs.} A condensation is continually fed material from the filament in which it is embedded. Within the condensation, the low angular momentum material rapidly contracts to form a protostar, and the remaining material orders itself into a protostellar disc and then slowly accretes onto the protostar, on a timescale determined by the strength of the gravitational torques which redistribute angular momentum within this disc. These protostellar accretion discs tend to be relatively compact and hot, and hence stable against fragmentation. Because accretion onto the discs is lumpy and irregular, they are continually disturbed. Consequently the discs do not settle down into a thin quasistatic equilibrium and there is strong vertical mixing.

{\sc tumbling filaments.} The angular momentum of an accretion disc is determined by the dynamics of the filament out of which it condenses. If the filament is tumbling (i.e. spinning about an axis at a significant angle to its length), the material flowing into the protostellar condensation at later times tends to have higher specific angular momentum, and as a consequence the disc will become quite extended, and may fragment. However, this is a rare occurrence, because the filaments do not often tumble significantly; their orientation in space tends to be approximately constant.

{\sc two representative examples.} In order to illustrate these properties, we describe in detail two runs from the ensemble of simulations we have performed, Runs 1b and 6d. We divide the detailed discussion into two section. Section 4 deals with the dynamics of filament formation and the formation of condensations. Section 5 deals with the dynamics inside a condensation, and in particular the properties of protostellar accretion discs.

{\sc the influence of the net angular momentum.} Although these simulations represent the extremes of the distribution of specific angular momenta, $j$ (see Table 1), we stress that $j$ is not the only -- nor even the main -- factor that leads them to follow distinctly different evolutionary paths. Nor is this due to different levels of turbulence, since the two simulations have similar values of $\gamma_{_{\rm TURB}}$. The divergence is mainly attributable to the stochastic details of the initial turbulent velocity fields. In Run 1b, the initial velocity field is mainly expansive, and so the primary protostar only forms after this expansion has been reversed and the material falls back on itself. The accretion disc also grows very slowly, so it is never sufficiently massive and/or extended to fragment. Conversely, in Run 6d the initial velocity field is compressive, and the primary protostar forms more quickly, although its condensation is then held up somewhat by rotation, and the subsequent growth of the primary protostar by accretion from the attendant protostellar disc is slow (much slower than in Run 1b).

\section{Global dynamics and filament formation}

{\sc The formation of filaments.} Figure \ref{FIG:FILAMENTS} shows false-colour column-density images on the principal cartesian planes for Runs 1b and 6d, with column-averaged velocity vectors superimposed. In Run 1b, (the three lefthand columns of Fig. \ref{FIG:FILAMENTS}), the initial velocity field has large positive $\nabla\cdot{\bf v}$ and this breaks the cloud into two pieces. When these two pieces turn around and fall back together, they form a filament, and the primary protostar condenses out towards one end of this filament. Since the filament is not tumbling significantly, the flux of material approaching the primary protostar carries little angular momentum, and it forms a rather compact hot disc, which is unable to fragment. By contrast, in Run 6d (the three righthand columns of Fig. \ref{FIG:FILAMENTS}), the initial velocity field is both compressive and rotational. As a result it quickly produces a tumbling filament, and then a protostar condenses out near the centre of the filament. Because the filament is tumbling, the fluxes of material approaching the protostar from opposite ends of the filament become offset from one another, and therefore deliver high angular momentum material into the condensation; consequently its disc is quite extended and prone to gravitational instabilities.

\begin{figure}
  \psfig{figure=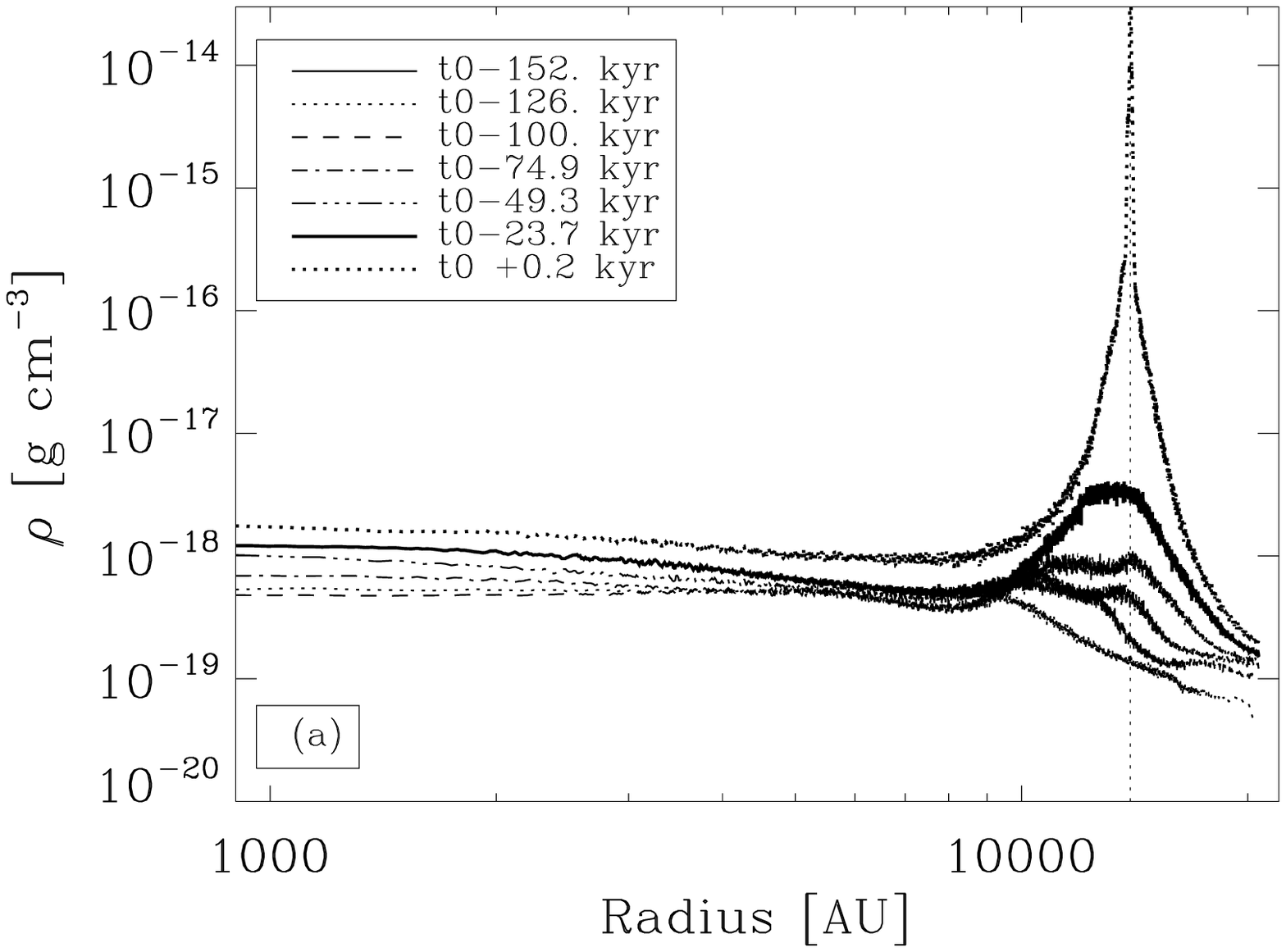 ,width=9cm}
  \psfig{figure=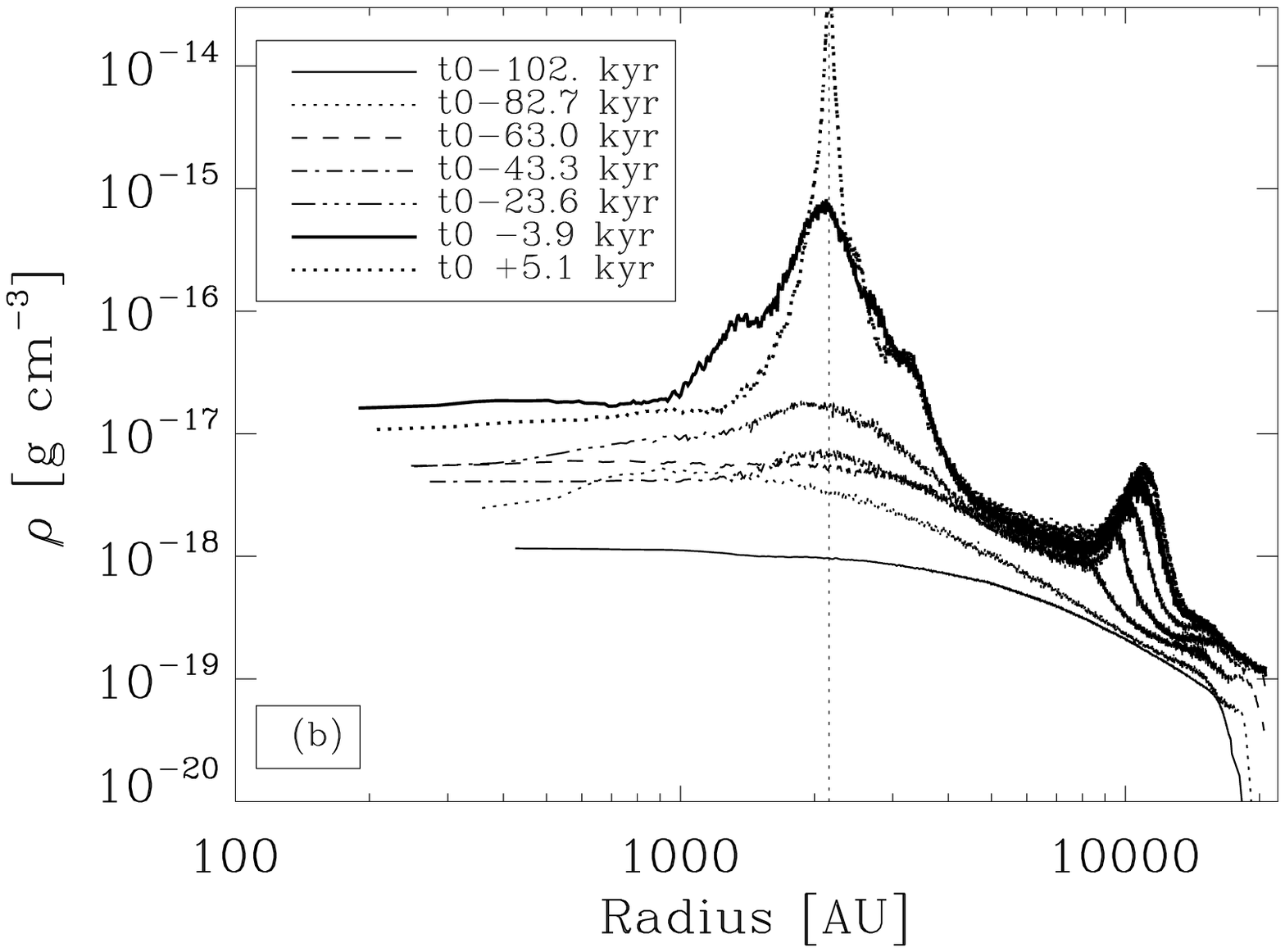 ,width=9cm}
\caption{The mean (spherically averaged) density at selected times. (a) Run 1b at $t-t_{_{\rm O}}=-152,\,-126,\,-100,\,-75,\,-49,\,-24\;{\rm and}\;0\;{\rm kyr}$. (b) Run 6d at $t-t_{_{\rm O}}=-102,\,-83,\,-63,\,-43,\,-24,\,-4\;{\rm and}\;+5\;{\rm kyr}$}\label{FIG:DENSITY}
\end{figure}

{\sc core density profiles.} Fig. \ref{FIG:DENSITY} shows $\bar{\rho}(r)$, the spherically averaged density profile, at various times from near the start of the simulation to just after the primary protostar forms at $t_{_{\rm O}}$. In Run 1b, the primary protostar condenses out near $r=15,000\,{\rm AU}$ (quite close to the edge of the initial Bonnor-Ebert sphere). In 6d, two condensations form out of the same filament, one at $r=2,000\,{\rm AU}$ that becomes the primary protostar, and one at $r=11,000\,{\rm AU}$; the latter may eventually become a wide companion, but the simulation had to be terminated before it had condensed out.

\begin{figure}
\psfig{figure=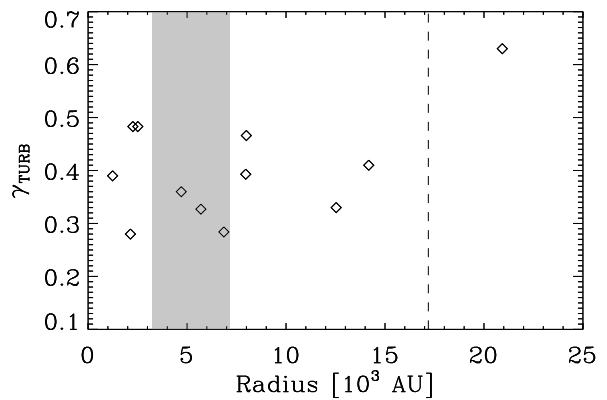, width=9cm}
\caption{Distances from the centre of the core at which primary protostars form vs. initial turbulent energy content $\gamma_{_{\rm TURB}}$ of the core. The shaded bar marks the transition region defined in the text, and the dashed line is the edge of the initial Bonnor-Ebert sphere. }\label{FIG:RFORM}
\end{figure}

{\sc the transition zone.} In the central zone of a Bonnor-Ebert sphere $-d\ln[\rho]/d\ln[r]\sim 0$, and in the outer zone $-d\ln[\rho]/d\ln[r]\sim 2$. We therefore define a transition zone where $0.5\leq -d\ln[\rho]/d\ln[r]\leq 1.5$. In terms of the dimensionless isothermal function \citep{Chandrasekhar1949}, the transition zone corresponds to $0.5\leq \xi\psi'\leq 1.5$, and hence $1.32\leq\xi\leq 2.90$. In our simulations, this translates into $3,250\,{\rm AU}\leq r\leq 7,150\,{\rm AU}$. Fig. \ref{FIG:RFORM} shows $\gamma_{_{\rm TURB}}$ plotted against the distance from the centre of the core at which each primary protostar forms; the shaded band is the transition zone defined above. Protostars are equally likely to form in the central zone, the transition zone, or the outer zone. Thus, the turbulent velocity field appears to erase quite effectively the initial density structure.

\begin{figure*}
  \epsfig{figure=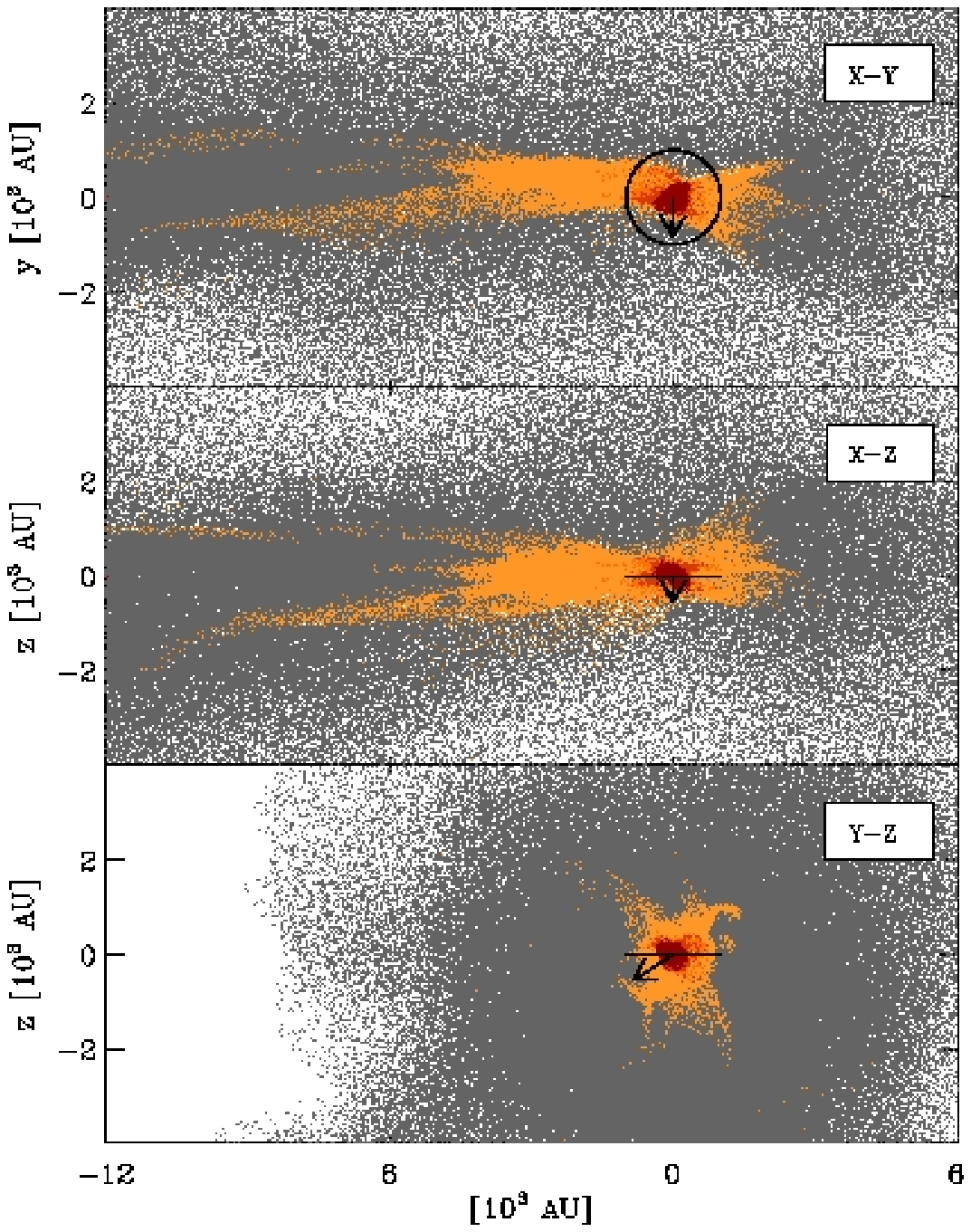, width=8.8cm}
  \epsfig{figure=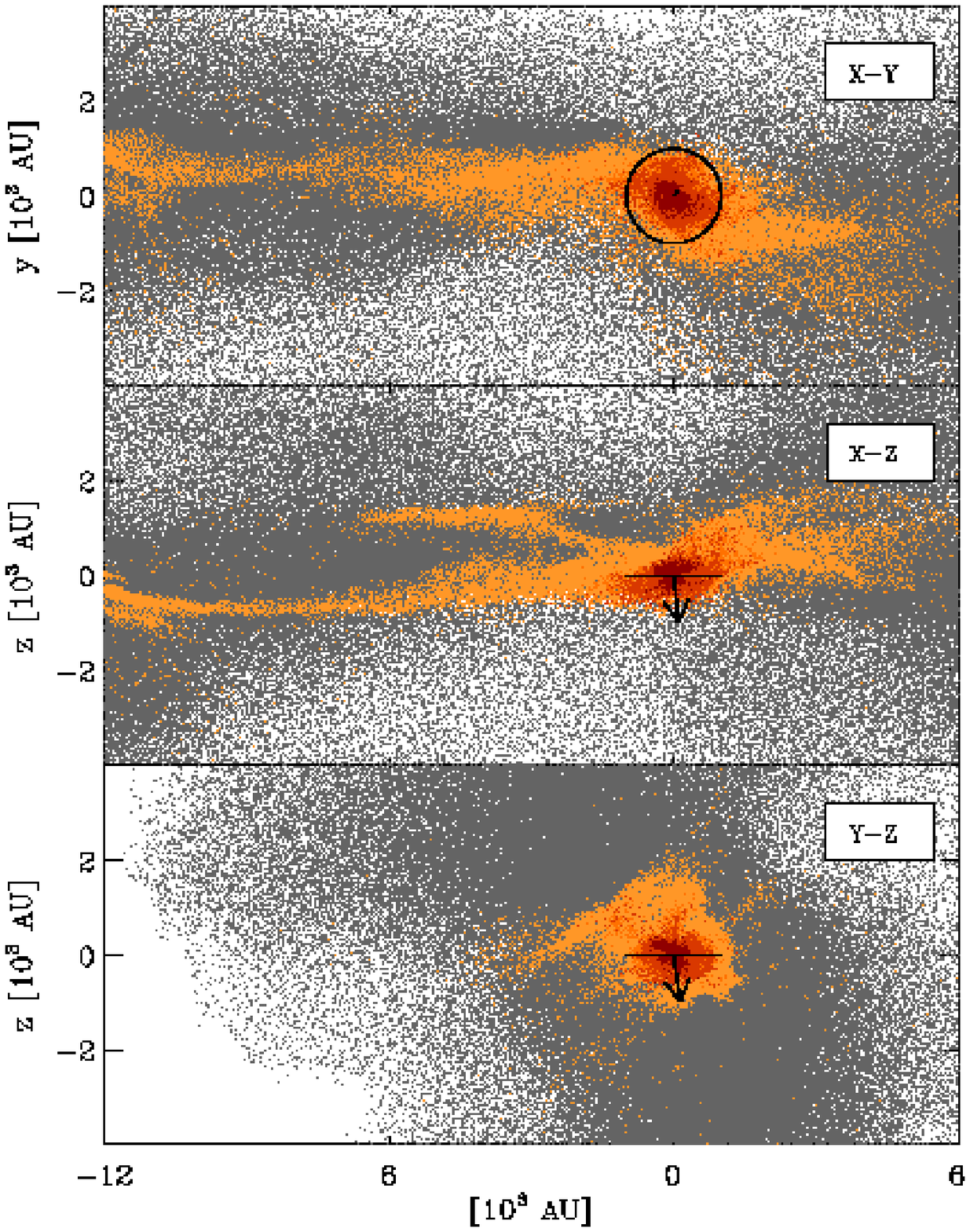, width=8.8cm}
\caption{SPH particles in the vicinity of the primary protostar in Run 1b at time $t=168\,{\rm kyr}$, and in Run 6d at time $t=138\,{\rm kyr}$. Particles are colour coded according as their temperature is $T<20\,{\rm K}$ (grey), $20\,{\rm K}\leq T\leq 50\,{\rm K}$ (orange), or $T>50\,{\rm K}$ (red). The co-ordinate system here is centered on the protostar and rotated according to the principal axes of the disc's system of inertia. The circle and the two straight lines indicate the projections in which the disc is seen, respectively, face on or edge-on; the disc itself is too small to be resolved on this plot. The arrows give the projected direction of the disc angular momentum. The arrows are normalised to $j=1.03 {\rm AU}^2 {\rm s}^{-1}$ for Run 1b, and to $j=0.19 {\rm AU}^2 {\rm s}^{-1}$ for Run 6d, respectively.}\label{FIG:PRECESSION}
\end{figure*}

{\sc the flow pattern near a condensation.} Fig. \ref{FIG:PRECESSION} shows the positions of SPH particles in the vicinity of the primary protostars is Runs 1b and 6d in the disc's system of inertia frame. The particles are colour coded according to their temperature. The dominant flow pattern involves material accreting onto the filament, then flowing along the filament towards and into the protostellar disc, and finally ending up in the protostar. As material accretes onto the filament it is heated to $\sim 20\,{\rm K}$ (this is evidenced by the sheath of warm particles at the edge of the filament), and thereafter it is heated further by compression, reaching $\sim 50\,{\rm K}$ as it approaches the disc.

{\sc the influence of filament dynamics on the condensation.} In Run 1b, the protostellar condensation is near the end of a filament, and so accretion onto the protostellar disc is quite lop-sided. In Run 6d, the protostellar condensation is close to the centre of a tumbling filament. Consequently comparable accretion flows converge on the condensation from either side. However, because the filament is tumbling, these flows become offset from one another (see Fig. \ref{FIG:FILAMENTS} and the $(x,y)$ projection in Fig. \ref{FIG:PRECESSION}). This is why the protostellar disc acquires a large amount of angular momentum, and is relatively extended.

\section{Disc properties}

\begin{figure*}
\label{FIG:DISCS}
  \begin{multicols}{2}
  \epsfig{figure=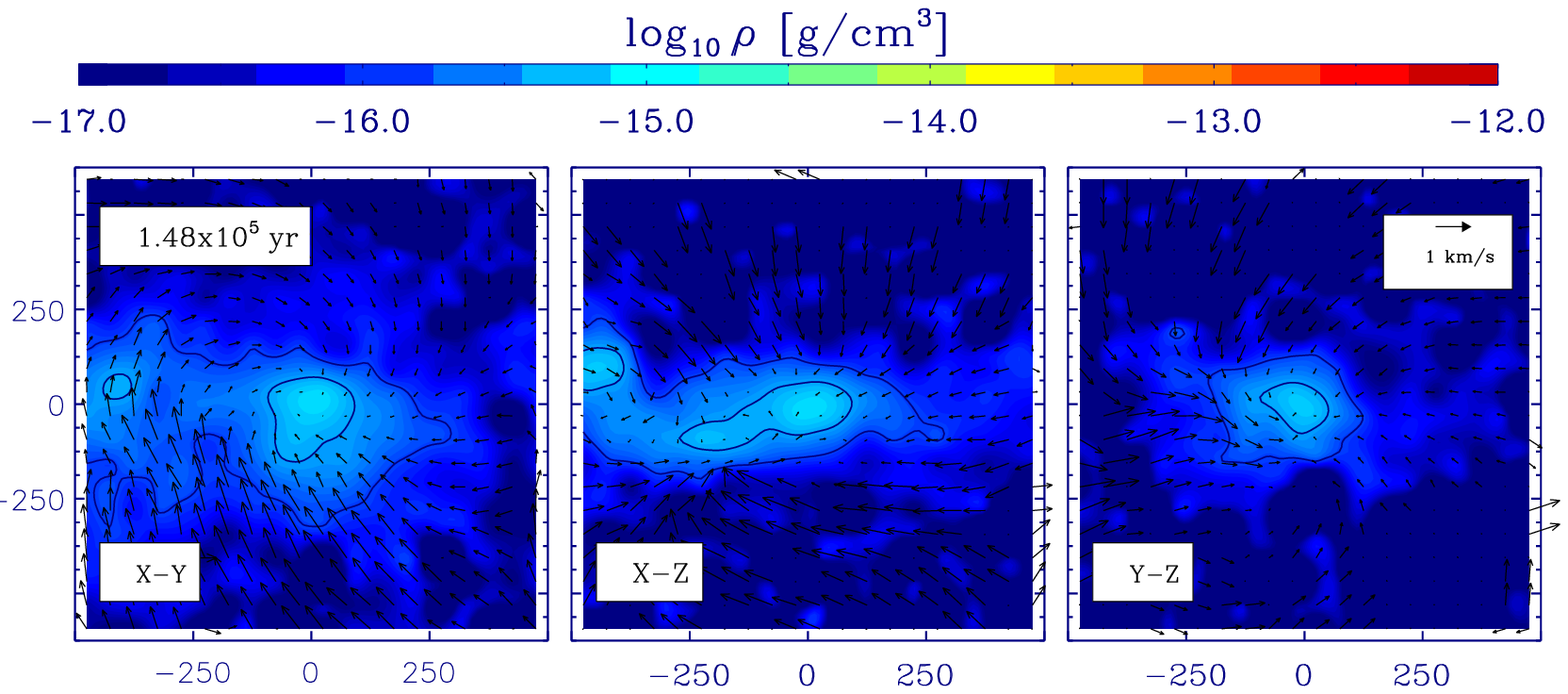 ,width=9.cm}
  \epsfig{figure=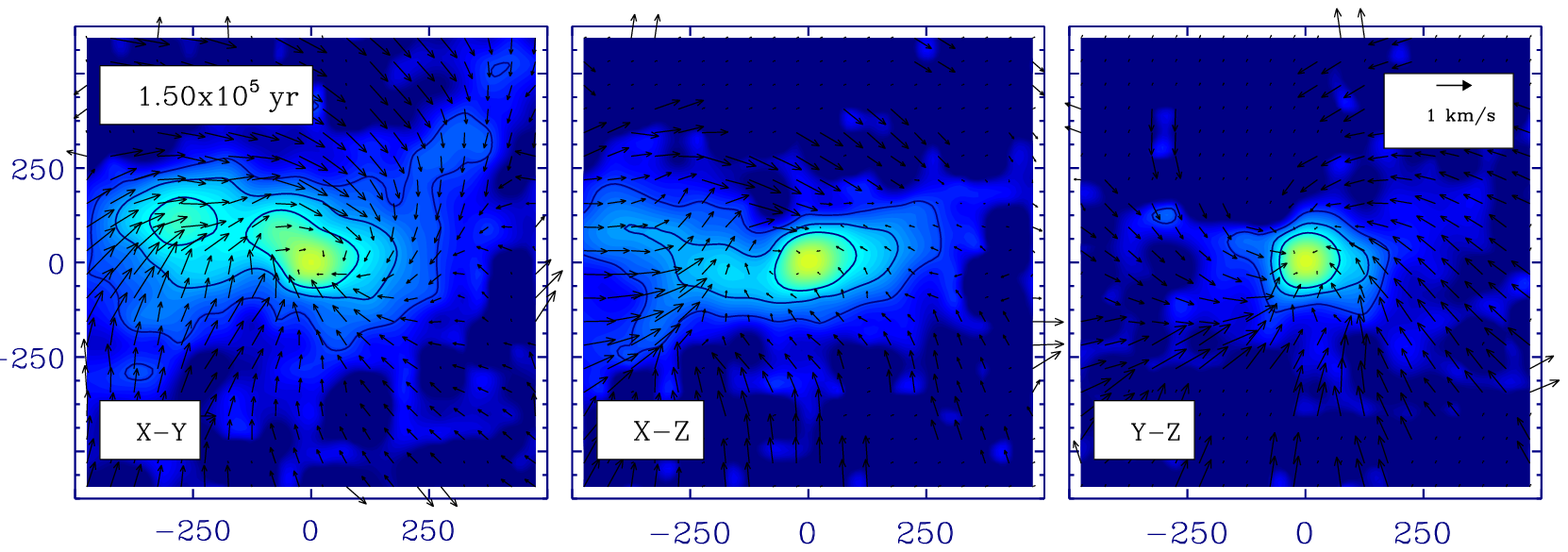 ,width=9.cm}
  \epsfig{figure=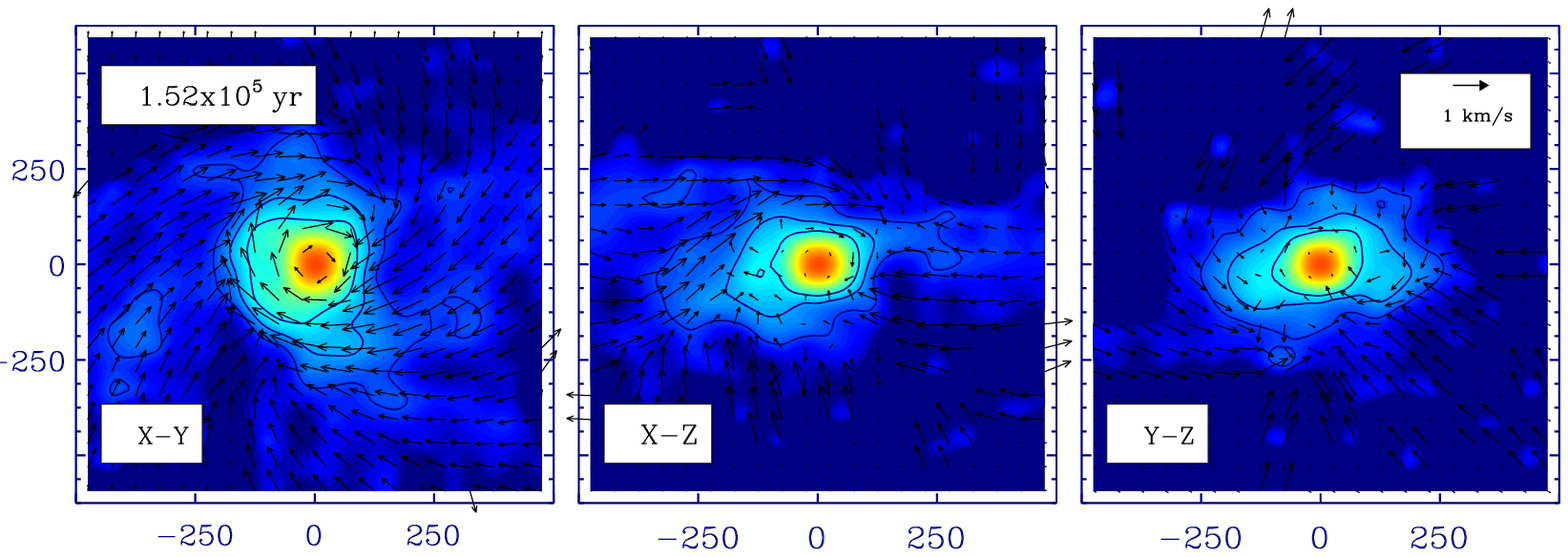 ,width=9.cm}
  \epsfig{figure=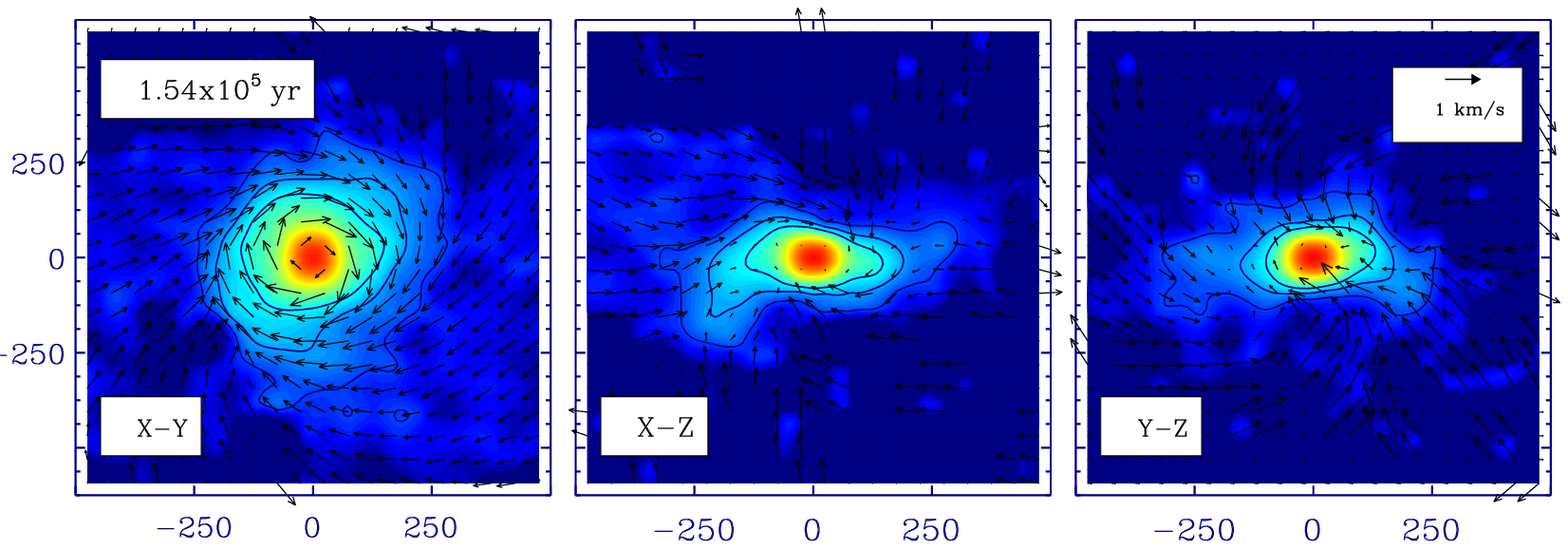 ,width=9.cm}
  \epsfig{figure=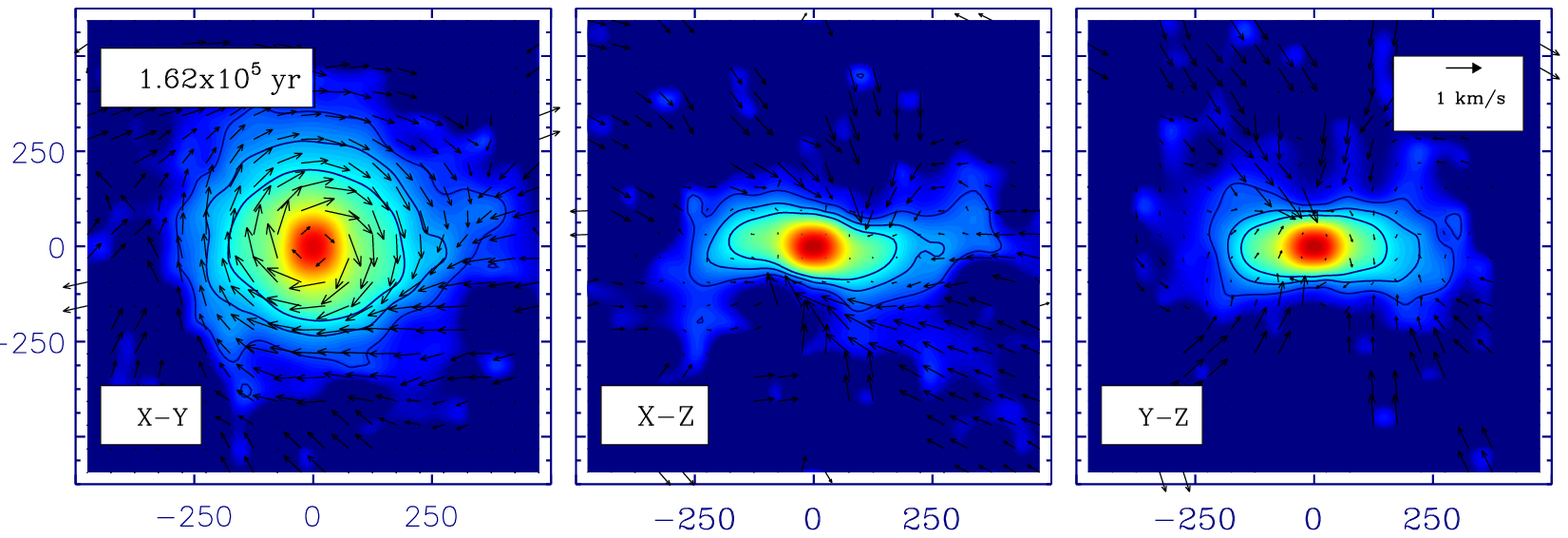 ,width=9.cm}
  \epsfig{figure=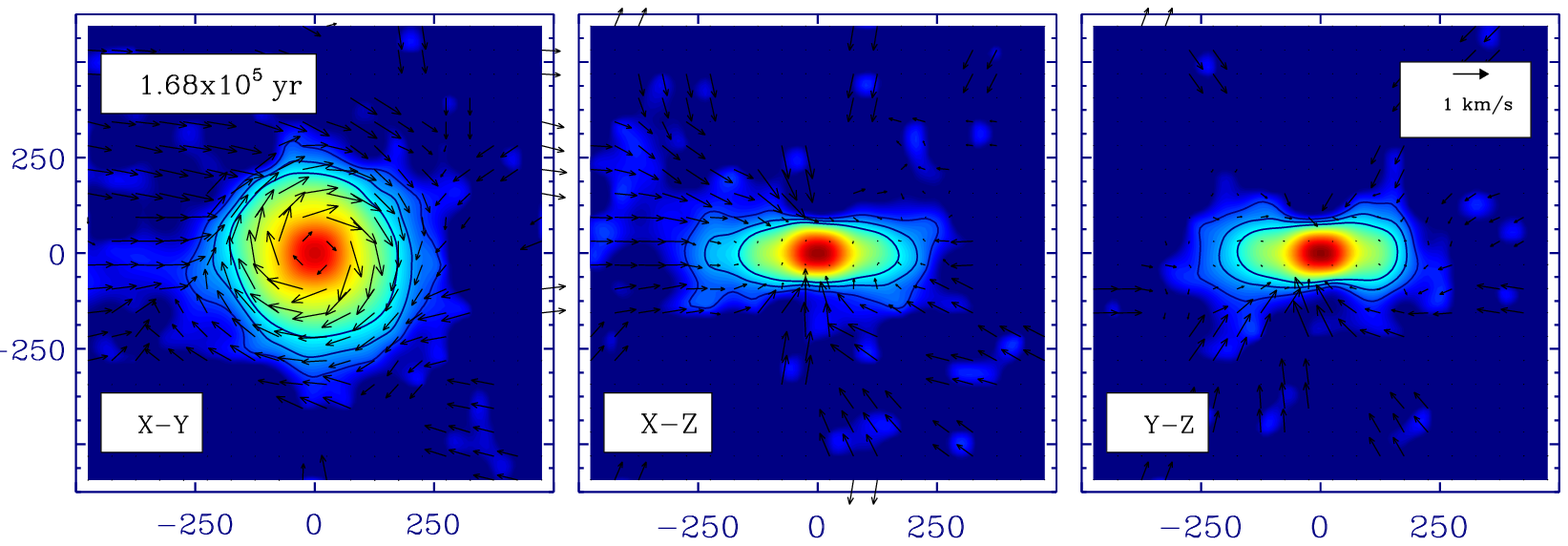 ,width=9.cm}\\

  \epsfig{figure=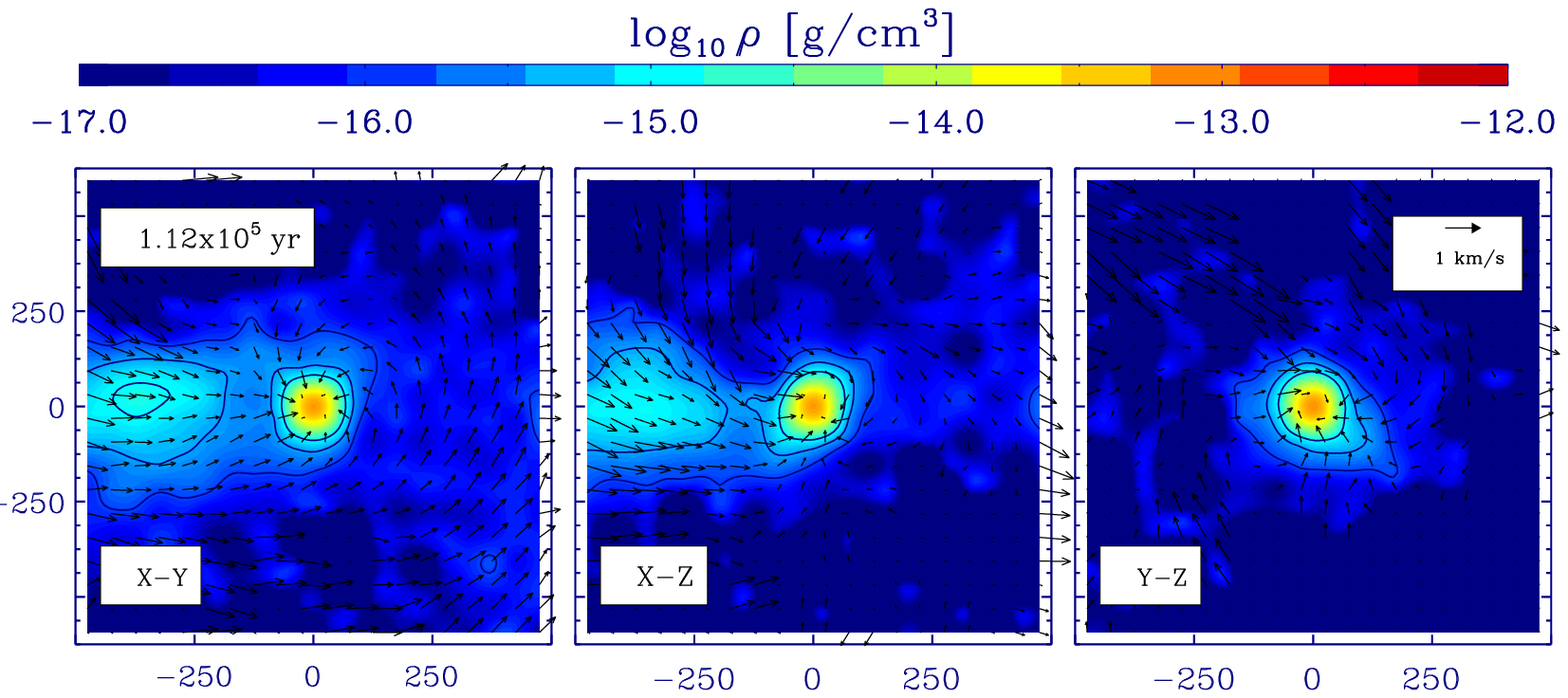 ,width=9.cm}
  \epsfig{figure=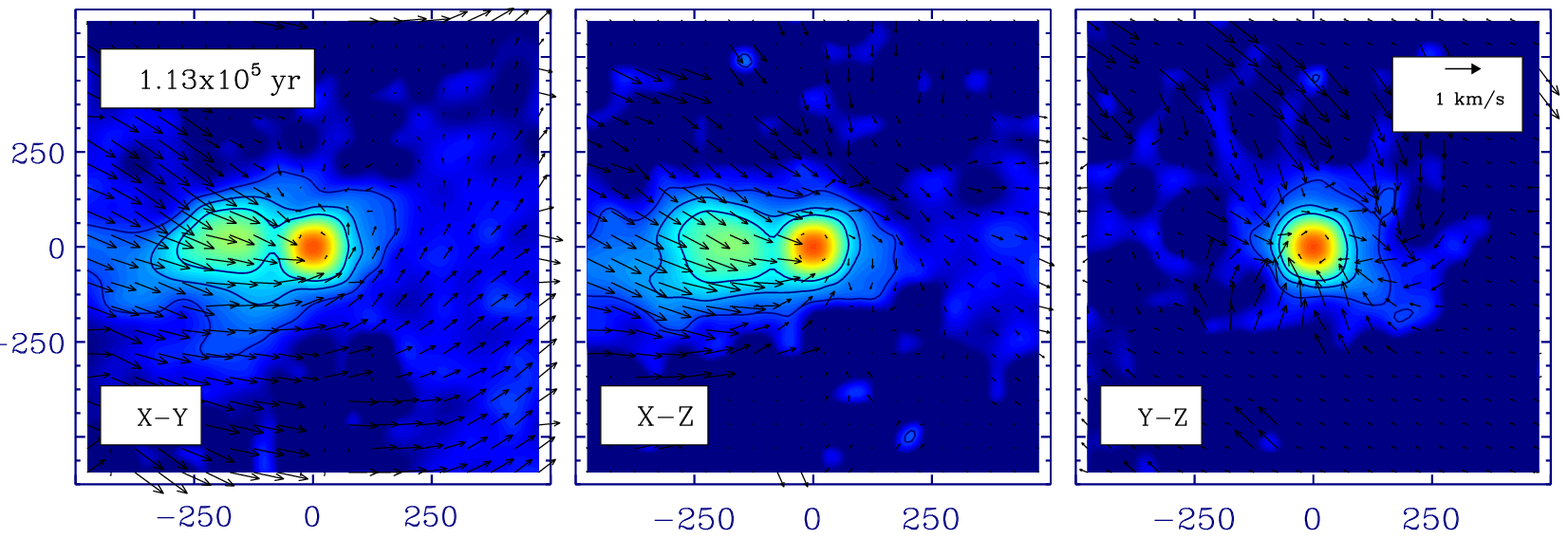 ,width=9.cm}
  \epsfig{figure=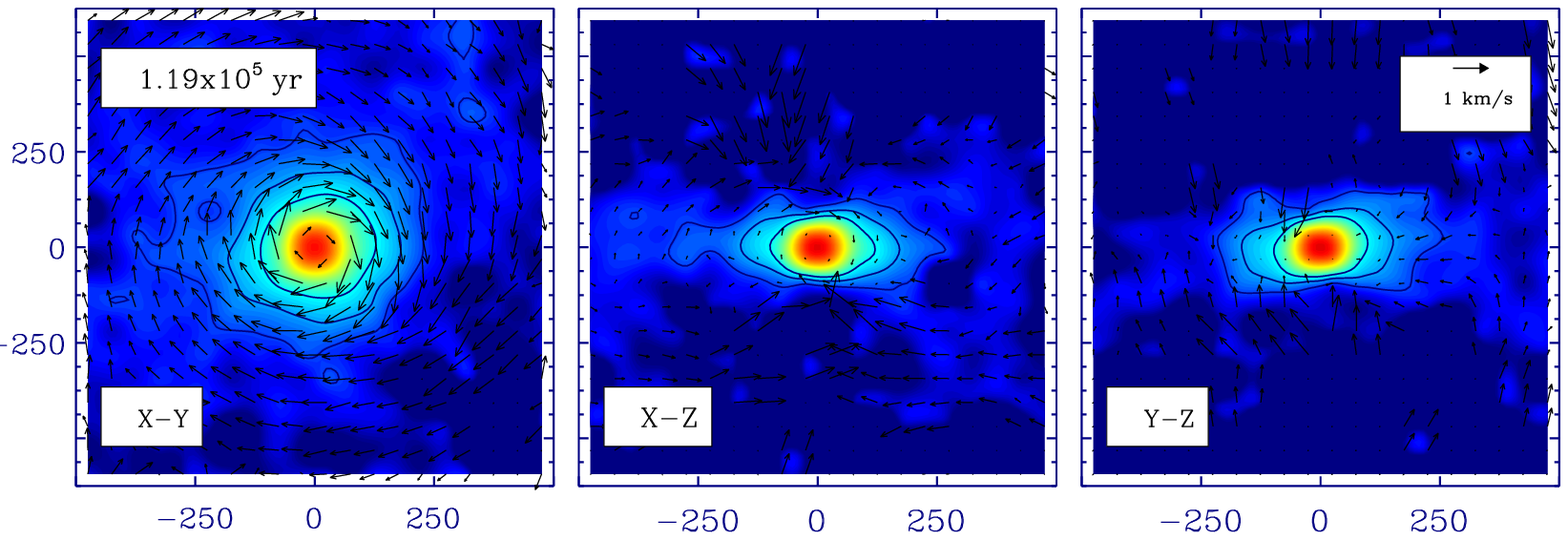 ,width=9cm}
  \epsfig{figure=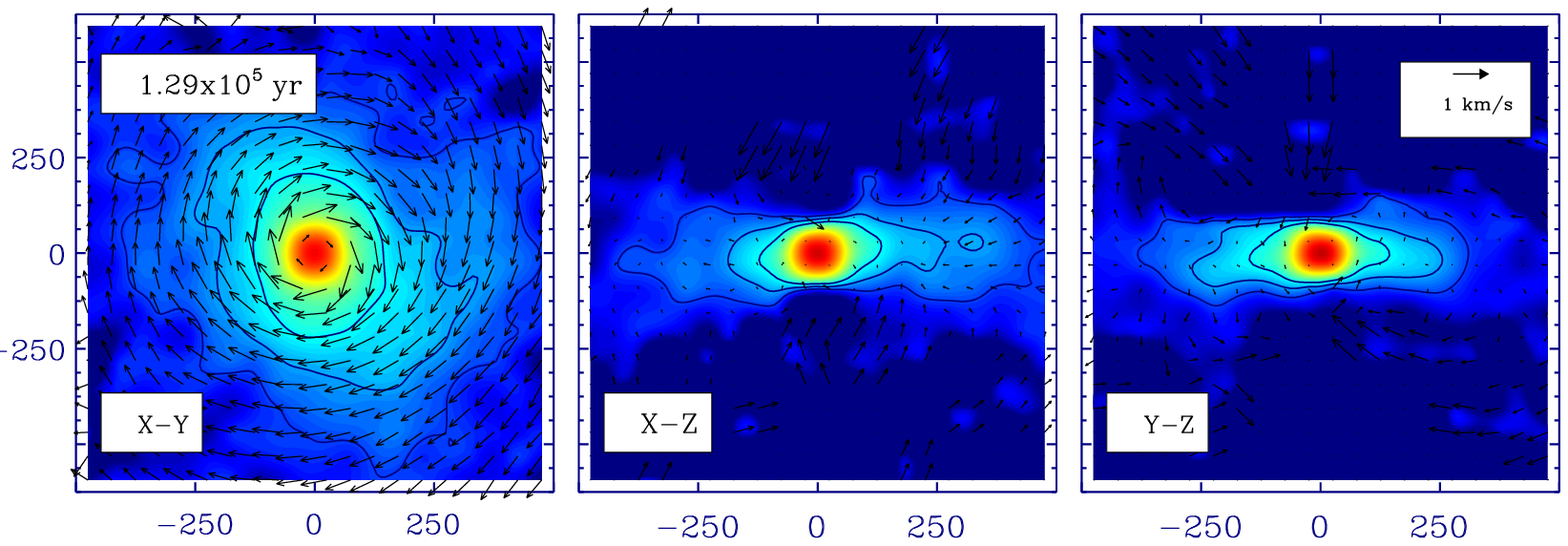 ,width=9cm}
  \epsfig{figure=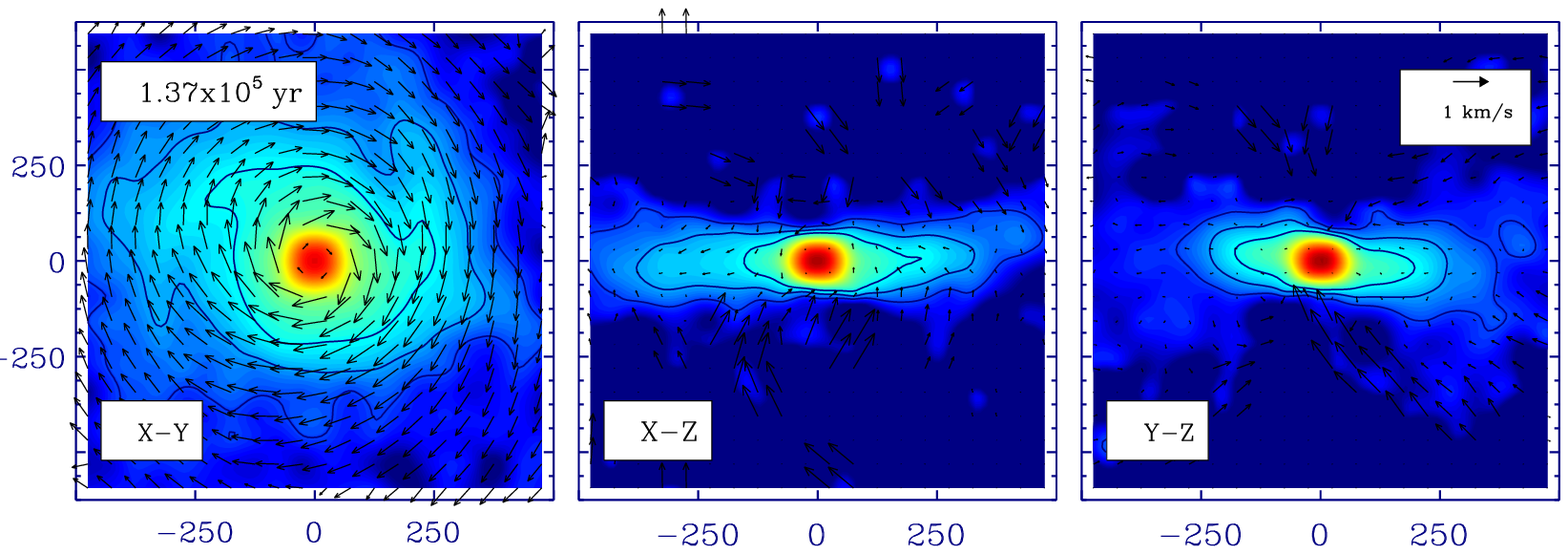 ,width=9cm}
  \epsfig{figure=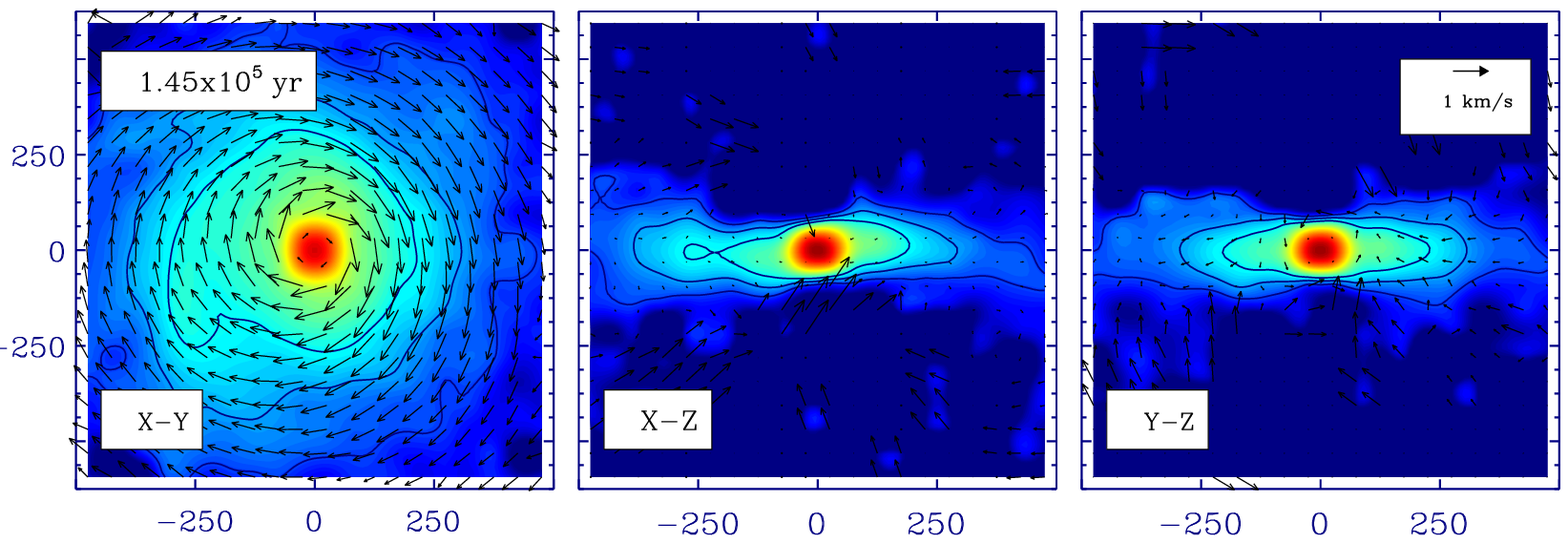 ,width=9cm}
\end{multicols}
\caption{False colour density slices on the principal cartesian planes through the protostellar discs formed in Runs 1b and 6d, with velocity vectors superimposed. The three lefthand columns show frames from Run 1b; successive rows correspond to the times $t=148,\,150,\,152,\,154,\,162\;{\rm and}\;168\;{\rm kyr}$. The three righthand columns show frames from Run 6d; successive rows correspond to the times $t=112,\,113,\,119,\,129,\,137\;{\rm and}\;145\;{\rm kyr}$. The density scale is shown at the top of the page, and a $1\,{\rm km}\,{\rm s}^{-1}$ velocity vector is given on the $x=0$ slices. All frames are $(1000\,{\rm AU})^2$, and the centre of co-ordinates has been shifted to the centre of mass of the disc.}
\end{figure*}

\begin{figure}
  \epsfig{figure=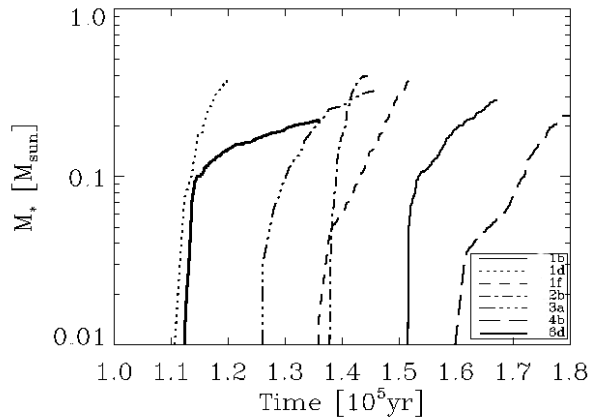,width=8cm}\label{mstars_turb}
\caption{The mass of the primary protostar as a function of time, for all the turbulent simulations.}\label{FIG:GROWTH}
\end{figure}

{\sc disc evolution.} Fig. \ref{FIG:DISCS} shows false-colour density images on the principal cartesian planes through the protostellar discs forming in Runs 1b and 6d. We see that the disc in Run 1b is much more compact and dense than that formed in Run 6d. Fig. \ref{FIG:GROWTH} shows the protostellar mass as a function of time, for all the turbulent simulations. We see that in Run 1b the protostar forms late ($t_{_{\rm O}}=150\,{\rm kyr}$), but then grows quite fast ($\dot{M}_\star\sim 10^{-5}\,{\rm M}_\odot\,{\rm yr}^{-1}$); this is typical of the majority of runs, in that the filament is not tumbling significantly and therefore material flows rapidly along the filament and either directly into the protostar or onto a compact protostellar disc. Conversely, in Run 6d the protostar forms early ($t_{_{\rm O}}=112\,{\rm kyr}$), but then grows more slowly ($\dot{M}_\star\sim 5\times 10^{-6}\,{\rm M}_\odot\,{\rm yr}^{-1}$); this is typical of a minority of runs, in that the filament is tumbling and therefore material flows initially into an extended protstellar discs and then spirals slowly onto the protostar.

{\sc termination of the simulations.} In W09 we followed all simulations until $28\%$ of the core mass was in the primary protostar or the protostellar disc. However, here the primary protostar -- once formed -- tends to grow more quickly, thereby slowing down the simulation dramatically; basically most of the processing power is being used following the motion of the dense material in the protostar. As a consequence we have had to terminate the simulations at arbitrary times, $t_{\rm END}$, when the evolution has become intolerably slow (see Table 1).

\begin{figure*}
  \psfig{figure=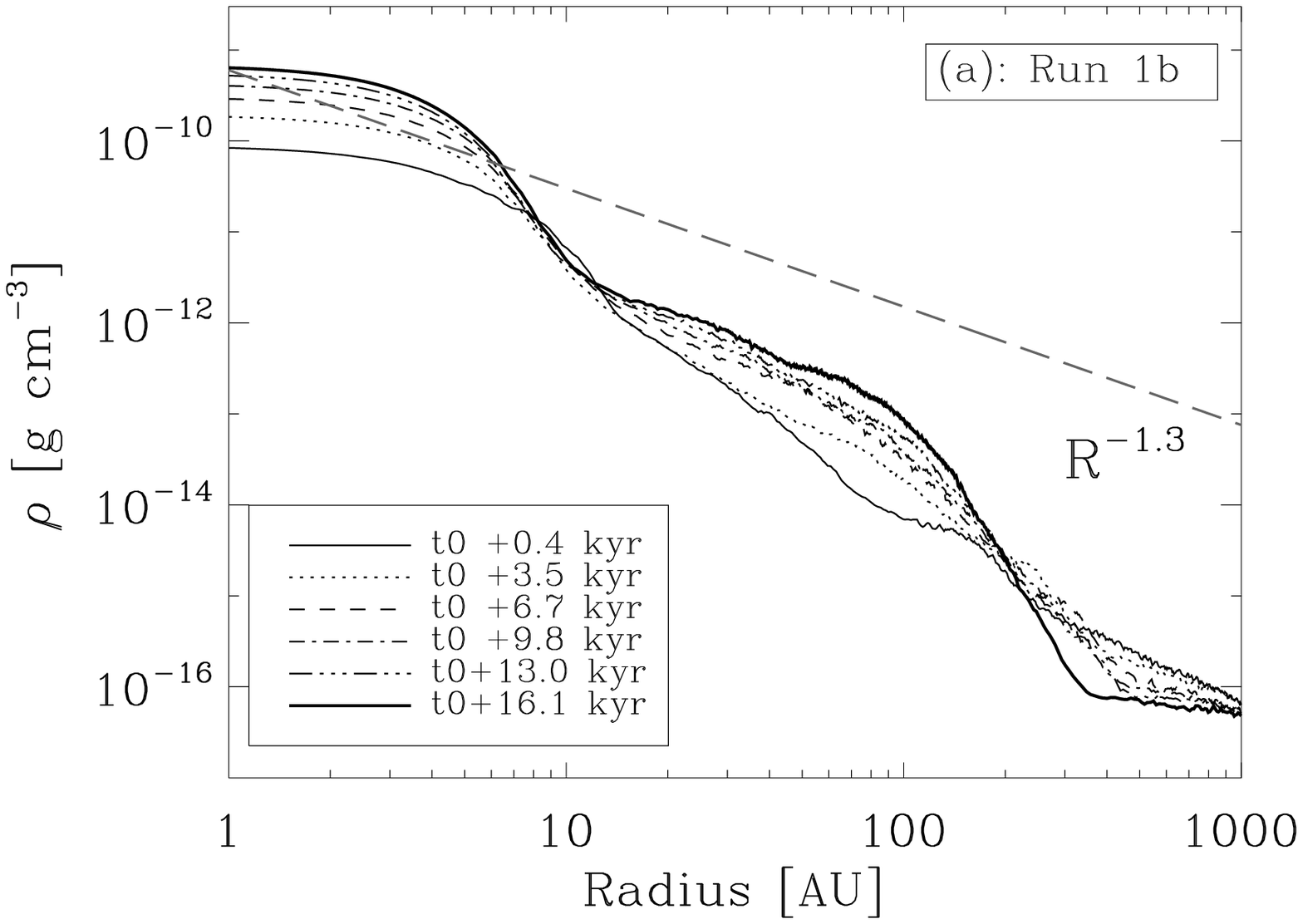, width=7cm}
  \psfig{figure=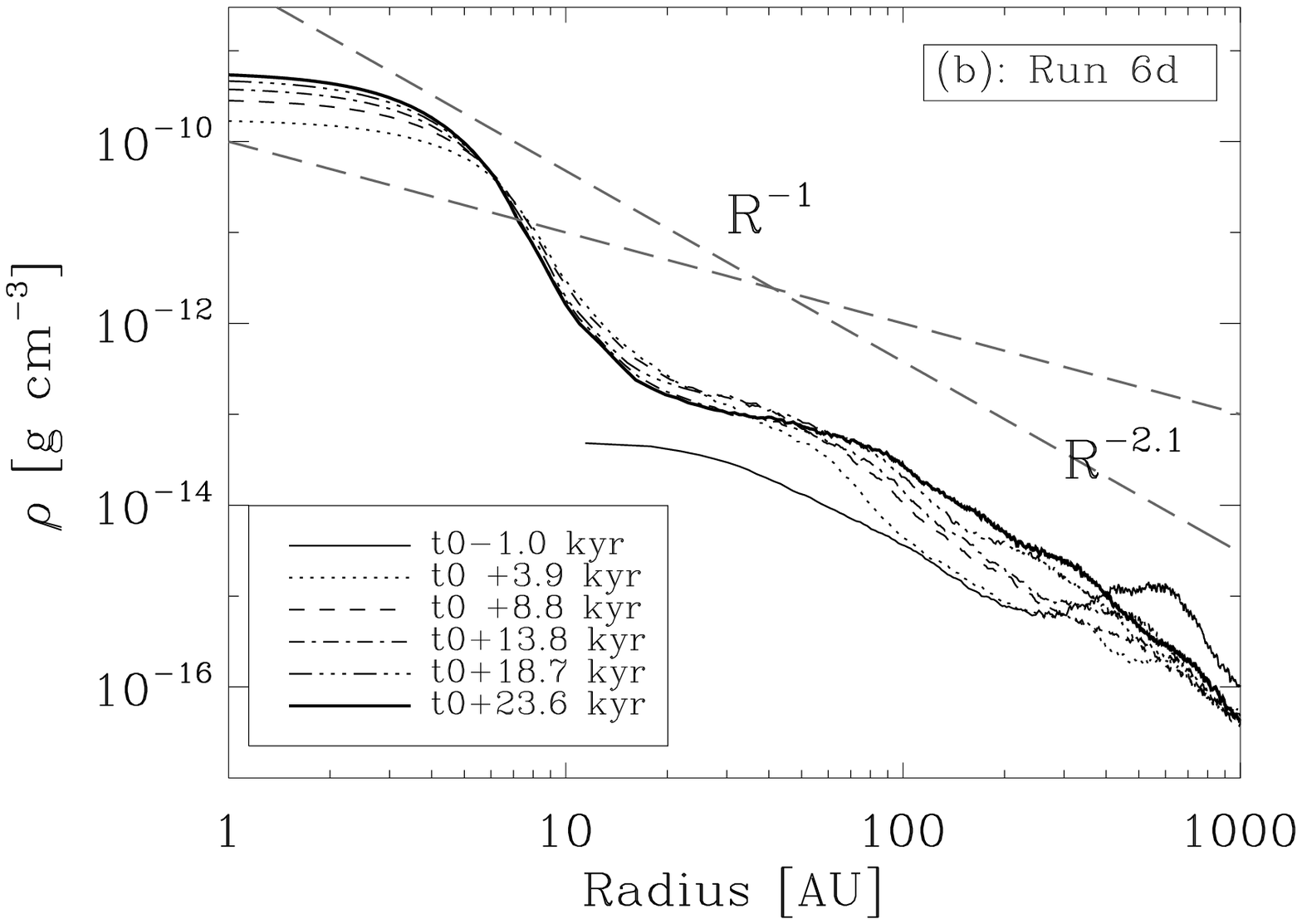, width=7cm}
  \psfig{figure=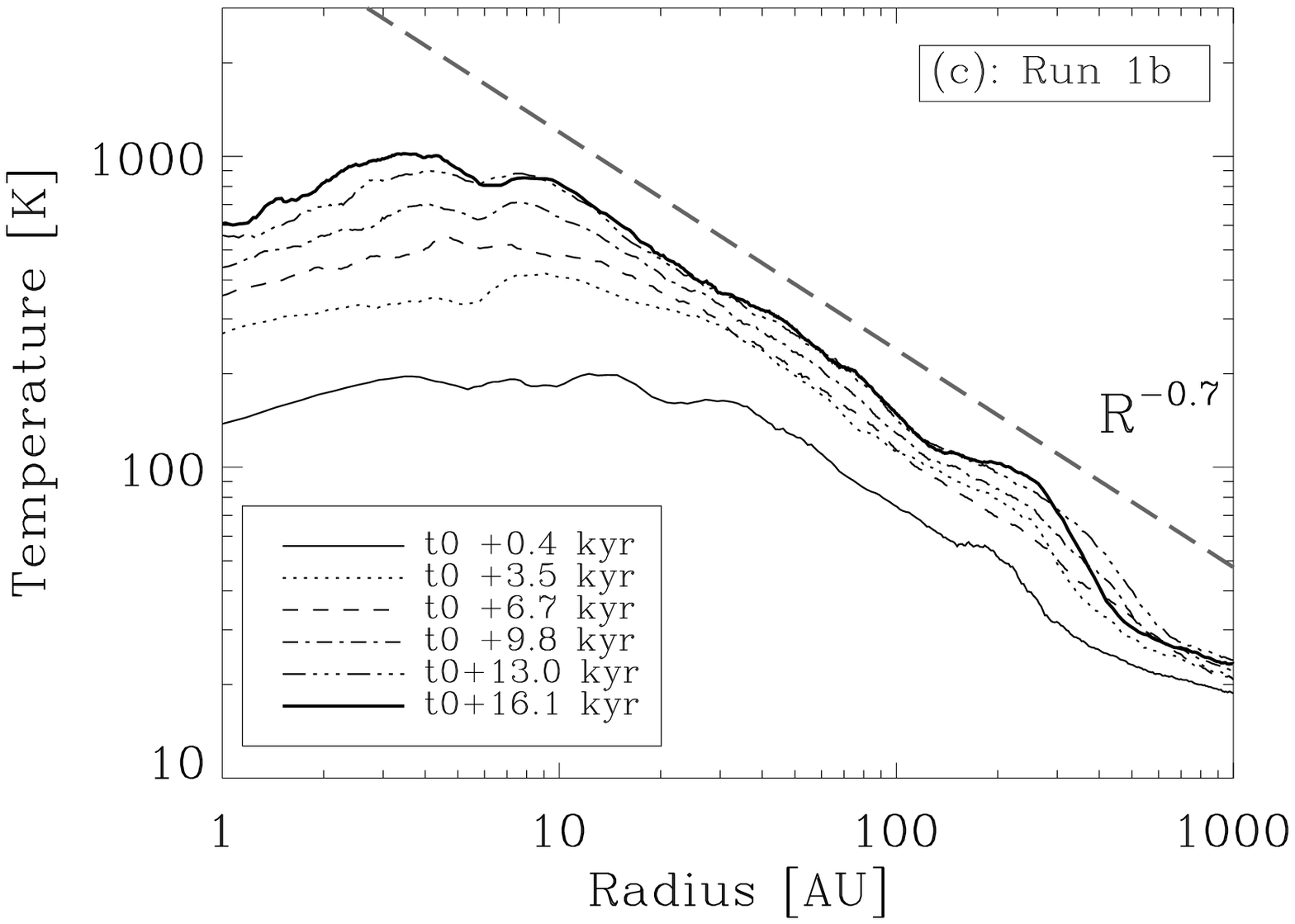, width=7cm}
  \psfig{figure=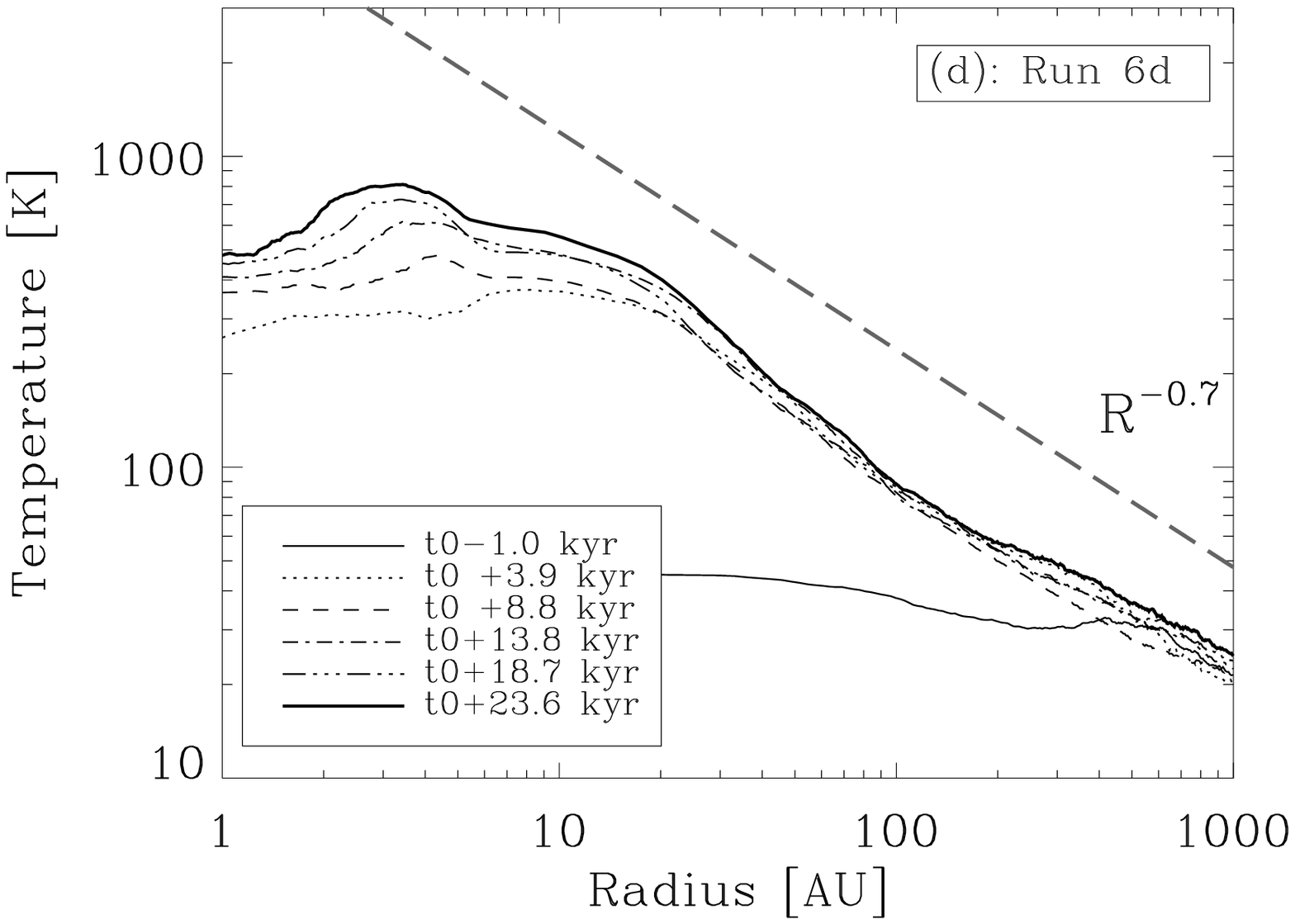, width=7cm}
\caption{Density and temperature profiles on the equatorial planes of the protostellar discs formed in Runs 1b and 6d. (a) $\rho(\omega,z=0)$ for Run 1b. (b) $\rho(\omega,z=0)$ for Run 6d. (c) $T(\omega,z=0)$ for Run 1b. (d) $T(\omega,z=0)$ for Run 6d. The centre of co-ordinates has been shifted to the centre of mass of the disc.}\label{FIG:PROFLES}
\end{figure*}

{\sc disc density profiles.} Figs. \ref{FIG:PROFLES}a,b show density profiles, $\rho(\omega,z=0)$, on the equatorial planes of the protostellar discs produced in Runs 1b and 6d. The disc in Run 1b has $\rho(\omega,z=0)\propto \omega^{-1.3}$, and extends out to $\sim 100\,{\rm AU}$. The disc in Run 6d has $\rho(\omega,z=0)\propto \omega^{-1}$, inside $\sim 100\,{\rm AU}$, and then the profile steepens to $\rho(\omega,z=0)\propto \omega^{-2.1}$ between $\sim 100\,{\rm AU}$ and $\sim 1000\,{\rm AU}$; it is more extended, and less dense, than the disc in Run 1b.

{\sc disc temperature profiles.} Figs. \ref{FIG:PROFLES}c,d show temperature profiles, $T(w,z=0)$, on the equatorial planes of the protostellar discs produced in Runs 1b and 6d. They both approximate to $T\propto w^{-0.7}$ (i.e. marginally steeper than for the discs formed from rigidly rotating cores in W09). The disc in Run 1b is somewhat hotter than that in Run 6d, because it is denser and more compact, and therefore cools less efficiently.

\begin{figure}
 \psfig{figure=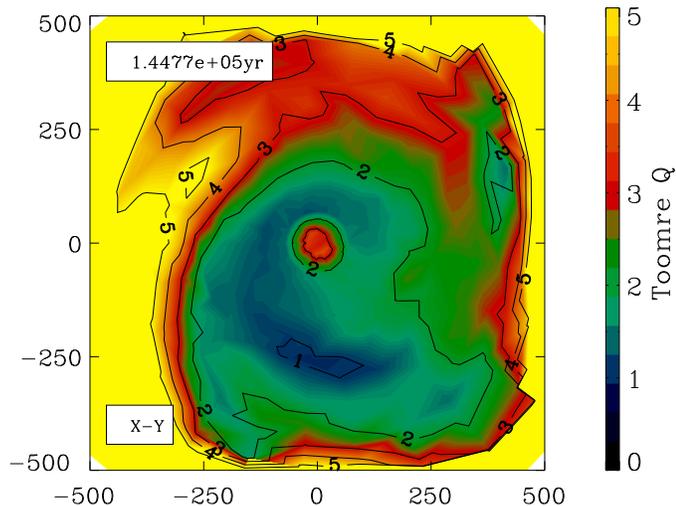 ,width=9cm}
\caption{False-colour image of the Toomre parameter, $Q_{_{\rm T}}$, for the disc formed in Run 6d, at time $t=144.8\,{\rm kyr}$. The centre of co-ordinates has been shifted to the centre of mass of the disc.}\label{FIG:STABILITY}
\end{figure}

{\sc disc stability.} Fig. \ref{FIG:STABILITY} shows the Toomre parameter, $Q_{_{\rm T}}$, for the relatively cool and extended disc formed in Run 6d. We see that there is a small unstable region in one of the spiral arms, and therefore it is possible that a fragment would condense out here, if the simulation could be followed further. All other discs seem to be gravitationally stable at $t_{_{\rm END}}$.

{\sc the origins of protostellar angular momentum.} Protostellar discs tend to be oriented with their rotation axes perpendicular to the filament in which they condense (see Fig. \ref{FIG:PRECESSION}). This implies that the net angular momentum of a filament derives from the fact that the filament is tumbling about one of its short axes, rather than spinning about its long axis  \citep[as, for example, in the model of ][]{Banerjee2006}. This agrees broadly with the finding of \citet{Anat2008}. who analysed the projected angles between filaments and the outflows from protostars embedded within the filaments. On the assumptions (a) that the outflow direction corresponds to the angular momentum of the underlying accretion disc, and (b) that their sample was randomly oriented and not subject to any selection effects, they inferred that most outflows are at a large angle ($\theta >45^{^{\rm o}}$) to the filament in which the driving protostar is embedded. However, we should emphasise that the observed filaments analysed by \citeauthor{Anat2008} are larger than the prestellar cores they contain, whereas the filaments formed here are within prestellar cores.

{\sc disc precession.} From Fig. \ref{FIG:PRECESSION} we see that the angular momenta of the protostellar discs are not in general aligned perpendicular to the instantaneous plane of the disc. This indicates that the discs are precessing, which is to be expected, because, in a turbulent environment, the disc is assembled by irregular, asymmetric and lumpy inflow, and the direction of its net angular momentum is repeatedly changing.

\section{Conclusions}
We have simulated star formation in prestellar cores with transonic turbulence. The cores are all modeled as supercritical Bonnor-Ebert spheres ($\xi_{_{\rm B}}=6.9$), with the density increased by $10\%$. All the cores have the same mass, $M_{_{\rm O}}=6.1\,{\rm M}_\odot$, and the same radius, $R_{_{\rm O}}=17,000\,{\rm AU}$, but different realisations of the initial turbulent velocity field. The simulated realisations have been selected from a large ensemble of four hundred different realisations, so as to span the same range of specific angular momenta, $j$, as the simulations performed in W09. In the rigidly rotating cores treated in W09, a collapse wave propagates from the outside in, and the density in the centre of the core increases monotonically until this collapse wave arrives after about one freefall time, $t_{_{\rm FF}}$. At that stage a protostar forms, surrounded by a protostellar disc. The extent of the disc depends on $j$, and only cores with large $j$ produce extended protostellar discs which are prone to fragment. In contrast, the key features of the turbulent cores simulated here are the following.

\begin{itemize}
\item{The outcome is not strongly dependent on $j$. Rather the outcome reflects the stochastic nature of the initial turbulent velocity field. The gas at the centre of the core is perturbed from the outset, and turbulent flows tend to sweep up the gas of the core into a filament, whose extent is comparable with the diameter of the core ($\sim 30,000\,{\rm AU}$).}
\item{Protostars condense out of these filaments. If two proto-fragments form in a filament at widely separated locations, they may survive to form a wide binary system, but usually proto-fragments in the same filament tend to merge before they can condense out. Proto-fragments can form at locations significantly displaced from the centre of the core, and they form with peculiar velocities of order $0.1\,{\rm km}\,{\rm s}^{-1}$.}
\item{Proto-fragments condense to form a protostar and then an attendant protostellar disc. However, these protostellar discs tend to be quite compact, with a shallow density profile, and stable against gravitational fragmentation or the formation of spiral waves. Part of the reason why they are stable appears to derive from the extra dynamical heating they experience as a result of lumpy, irregular accretion from the turbulent core envelope.}
\item{The intrinsic angular momentum of a proto-stellar disc, and hence also its rotation axis, tend to be perpendicular to the long axis of the birth filament. This suggests that the angular momentum of a proto-fragment derives from the fact that the filaments are tumbling, rather than spinning about their long-axis.}
\item{A key parameter influencing the statistics of turbulence in a core is the minimum wavenumber, $k_{_{\rm MIN}}$. Only if $k_{_{\rm MIN}}=1$ (i.e. turbulence is injected on the scale of the diameter of the core) do we reproduce the range of specific angular momenta observed in nature. Thus it is the largest wavelengths in the turbulent spectrum which are critical. This accords with the result reported recently by \citet{Bate2009b}, who shows that the statistics of stars formed in clouds with $P_k\propto k^{-4}$ and $P_k\propto k^{-6}$ are indistinguishable.}
\end{itemize}

\section*{Acknowledgments}
We thank D. Neufeld for providing the cooling tables. S. Walch performed this work with support from the International Max-Planck Research School, and the DFG Cluster of Excellence {\it Origin and Structure of the Universe} (www.universe-cluster.de). SW and AW gratefully acknowledge the support of the the EC-funded Marie Curie Research Training Network {\sc constellation} (MRTN-CT-2006-035890). 

\bibliographystyle{mn2e}
\bibliography{references}

\begin{thebibliography}{}

\bibitem[\protect\citeauthoryear{{Alves}, {Lombardi} \& {Lada}}{{Alves}
  et~al.}{2007}]{Alves2007}
{Alves} J.,  {Lombardi} M.,    {Lada} C.~J.,  2007, \aap, 462, L17

\bibitem[\protect\citeauthoryear{{Anathpindika} \& {Whitworth}}{{Anathpindika}
  \& {Whitworth}}{2008}]{Anat2008}
{Anathpindika} S.,  {Whitworth} A.~P.,  2008, \aap, 487, 605

\bibitem[\protect\citeauthoryear{{Andre}, {Ward-Thompson} \& {Barsony}}{{Andre}
  et~al.}{1993}]{Andre1993}
{Andre} P.,  {Ward-Thompson} D.,    {Barsony} M.,  1993, \apj, 406, 122

\bibitem[\protect\citeauthoryear{{Attwood}, {Goodwin}, {Stamatellos} \&
  {Whitworth}}{{Attwood} et~al.}{2009}]{Attwood2009}
{Attwood} R.~E.,  {Goodwin} S.~P.,  {Stamatellos} D.,    {Whitworth} A.~P.,
  2009, \aap, 495, 201

\bibitem[\protect\citeauthoryear{{Balsara}}{{Balsara}}{1995}]{Balsara1995}
{Balsara} D.~S.,  1995, Journal of Computational Physics, 121, 357

\bibitem[\protect\citeauthoryear{{Banerjee}, {Pudritz} \&
  {Anderson}}{{Banerjee} et~al.}{2006}]{Banerjee2006}
{Banerjee} R.,  {Pudritz} R.~E.,    {Anderson} D.~W.,  2006, \mnras, 373, 1091

\bibitem[\protect\citeauthoryear{{Banerjee}, {Pudritz} \& {Holmes}}{{Banerjee}
  et~al.}{2004}]{Banerjee2004}
{Banerjee} R.,  {Pudritz} R.~E.,    {Holmes} L.,  2004, \mnras, 355, 248

\bibitem[\protect\citeauthoryear{{Barranco} \& {Goodman}}{{Barranco} \&
  {Goodman}}{1998}]{Barranco1998}
{Barranco} J.~A.,  {Goodman} A.~A.,  1998, \apj, 504, 207

\bibitem[\protect\citeauthoryear{{Bate}}{{Bate}}{2009a}]{Bate2009}
{Bate} M.~R.,  2009a, \mnras, 392, 590

\bibitem[\protect\citeauthoryear{{Bate}}{{Bate}}{2009b}]{Bate2009b}
{Bate} M.~R.,  2009b, ArXiv e-prints

\bibitem[\protect\citeauthoryear{{Bate} \& {Burkert}}{{Bate} \&
  {Burkert}}{1997}]{BateBurkert97}
{Bate} M.~R.,  {Burkert} A.,  1997, \mnras, 288, 1060

\bibitem[\protect\citeauthoryear{{Beichman}, {Myers}, {Emerson}, {Harris},
  {Mathieu}, {Benson} \& {Jennings}}{{Beichman} et~al.}{1986}]{Beichman1986}
{Beichman} C.~A.,  {Myers} P.~C.,  {Emerson} J.~P.,  {Harris} S.,  {Mathieu}
  R.,  {Benson} P.~J.,    {Jennings} R.~E.,  1986, \apj, 307, 337

\bibitem[\protect\citeauthoryear{{Bonnell}, {Clark} \& {Bate}}{{Bonnell}
  et~al.}{2008}]{Bonnell2008}
{Bonnell} I.~A.,  {Clark} P.,    {Bate} M.~R.,  2008, \mnras, 389, 1556

\bibitem[\protect\citeauthoryear{{Burkert} \& {Bodenheimer}}{{Burkert} \&
  {Bodenheimer}}{2000}]{BurkertBodenheimer00}
{Burkert} A.,  {Bodenheimer} P.,  2000, \apj, 543, 822

\bibitem[\protect\citeauthoryear{{Burkert} \& {Hartmann}}{{Burkert} \&
  {Hartmann}}{2004}]{Burkert2004}
{Burkert} A.,  {Hartmann} L.,  2004, \apj, 616, 288

\bibitem[\protect\citeauthoryear{{Caselli}, {Benson}, {Myers} \&
  {Tafalla}}{{Caselli} et~al.}{2002}]{Caselli2002}
{Caselli} P.,  {Benson} P.~J.,  {Myers} P.~C.,    {Tafalla} M.,  2002, \apj,
  572, 238

\bibitem[\protect\citeauthoryear{{Chabrier}}{{Chabrier}}{2003}]{Chabrier2003}
{Chabrier} G.,  2003, \pasp, 115, 763

\bibitem[\protect\citeauthoryear{{Chandrasekhar} \& {Wares}}{{Chandrasekhar} \&
  {Wares}}{1949}]{Chandrasekhar1949}
{Chandrasekhar} S.,  {Wares} G.~W.,  1949, \apj, 109, 551

\bibitem[\protect\citeauthoryear{{Dubinski}, {Narayan} \&
  {Phillips}}{{Dubinski} et~al.}{1995}]{Dubinski1995}
{Dubinski} J.,  {Narayan} R.,    {Phillips} T.~G.,  1995, \apj, 448, 226

\bibitem[\protect\citeauthoryear{{Elmegreen}}{{Elmegreen}}{2000}]{Elmegreen200%
0}
{Elmegreen} B.~G.,  2000, \apj, 530, 277

\bibitem[\protect\citeauthoryear{{Gingold} \& {Monaghan}}{{Gingold} \&
  {Monaghan}}{1983}]{Gingold1983}
{Gingold} R.~A.,  {Monaghan} J.~J.,  1983, \mnras, 204, 715

\bibitem[\protect\citeauthoryear{{Goldsmith} \& {Arquilla}}{{Goldsmith} \&
  {Arquilla}}{1985}]{Goldsmith1985}
{Goldsmith} P.~F.,  {Arquilla} R.,  1985, in {Black} D.~C.,  {Matthews} M.~S.,
  eds, Protostars and Planets II {Rotation in dark clouds}.
pp 137--149

\bibitem[\protect\citeauthoryear{{Goodman}, {Barranco}, {Wilner} \&
  {Heyer}}{{Goodman} et~al.}{1998}]{Goodman1998}
{Goodman} A.~A.,  {Barranco} J.~A.,  {Wilner} D.~J.,    {Heyer} M.~H.,  1998,
  \apj, 504, 223

\bibitem[\protect\citeauthoryear{{Goodman}, {Benson}, {Fuller} \&
  {Myers}}{{Goodman} et~al.}{1993}]{Goodman1993}
{Goodman} A.~A.,  {Benson} P.~J.,  {Fuller} G.~A.,    {Myers} P.~C.,  1993,
  \apj, 406, 528

\bibitem[\protect\citeauthoryear{{Goodwin}, {Whitworth} \&
  {Ward-Thompson}}{{Goodwin} et~al.}{2004a}]{Goodwin2004}
{Goodwin} S.~P.,  {Whitworth} A.~P.,    {Ward-Thompson} D.,  2004a, \aap, 414,
  633

\bibitem[\protect\citeauthoryear{{Goodwin}, {Whitworth} \&
  {Ward-Thompson}}{{Goodwin} et~al.}{2004b}]{Goodwin2004a}
{Goodwin} S.~P.,  {Whitworth} A.~P.,    {Ward-Thompson} D.,  2004b, \aap, 414,
  633

\bibitem[\protect\citeauthoryear{{Goodwin}, {Whitworth} \&
  {Ward-Thompson}}{{Goodwin} et~al.}{2004c}]{Goodwin2004b}
{Goodwin} S.~P.,  {Whitworth} A.~P.,    {Ward-Thompson} D.,  2004c, \aap, 423,
  169

\bibitem[\protect\citeauthoryear{{Goodwin}, {Whitworth} \&
  {Ward-Thompson}}{{Goodwin} et~al.}{2006}]{Goodwin2006}
{Goodwin} S.~P.,  {Whitworth} A.~P.,    {Ward-Thompson} D.,  2006, \aap, 452,
  487

\bibitem[\protect\citeauthoryear{{Jijina}, {Myers} \& {Adams}}{{Jijina}
  et~al.}{1999}]{Jijina1999}
{Jijina} J.,  {Myers} P.~C.,    {Adams} F.~C.,  1999, \apjs, 125, 161

\bibitem[\protect\citeauthoryear{{Johnstone}, {Fich}, {Mitchell} \&
  {Moriarty-Schieven}}{{Johnstone} et~al.}{2001}]{Johnstone2001}
{Johnstone} D.,  {Fich} M.,  {Mitchell} G.~F.,    {Moriarty-Schieven} G.,
  2001, \apj, 559, 307

\bibitem[\protect\citeauthoryear{{Kroupa}}{{Kroupa}}{2001}]{Kroupa2001a}
{Kroupa} P.,  2001, \mnras, 322, 231

\bibitem[\protect\citeauthoryear{{Larson}}{{Larson}}{1981}]{Larson1981}
{Larson} R.~B.,  1981, \mnras, 194, 809

\bibitem[\protect\citeauthoryear{{Motte}, {Andre} \& {Neri}}{{Motte}
  et~al.}{1998}]{Motte1998}
{Motte} F.,  {Andre} P.,    {Neri} R.,  1998, \aap, 336, 150

\bibitem[\protect\citeauthoryear{{Myers} \& {Gammie}}{{Myers} \&
  {Gammie}}{1999}]{Myers1999}
{Myers} P.~C.,  {Gammie} C.~F.,  1999, \apjl, 522, L141

\bibitem[\protect\citeauthoryear{{Myers}, {Ladd} \& {Fuller}}{{Myers}
  et~al.}{1991}]{Myers1991}
{Myers} P.~C.,  {Ladd} E.~F.,    {Fuller} G.~A.,  1991, \apjl, 372, L95

\bibitem[\protect\citeauthoryear{{Nelson}, {Wetzstein}, {Naab} \& {.}}{{Nelson}
  et~al.}{2008}]{Nelson2008}
{Nelson} A.~F.,  {Wetzstein} M.,  {Naab} T.,    {.} 2008, ArXiv e-prints, 802,
  0802.4253

\bibitem[\protect\citeauthoryear{{Neufeld}, {Lepp} \& {Melnick}}{{Neufeld}
  et~al.}{1995}]{Neufeld1995}
{Neufeld} D.~A.,  {Lepp} S.,    {Melnick} G.~J.,  1995, \apjs, 100, 132

\bibitem[\protect\citeauthoryear{{Nutter} \& {Ward-Thompson}}{{Nutter} \&
  {Ward-Thompson}}{2007}]{Nutter2007}
{Nutter} D.,  {Ward-Thompson} D.,  2007, \mnras, 374, 1413

\bibitem[\protect\citeauthoryear{{Offner}, {Klein} \& {McKee}}{{Offner}
  et~al.}{2008}]{Offner2008}
{Offner} S.~S.~R.,  {Klein} R.~I.,    {McKee} C.~F.,  2008, \apj, 686, 1174

\bibitem[\protect\citeauthoryear{{Offner}, {Klein}, {McKee} \&
  {Krumholz}}{{Offner} et~al.}{2009}]{Offner2009}
{Offner} S.~S.~R.,  {Klein} R.~I.,  {McKee} C.~F.,    {Krumholz} M.~R.,  2009,
  ArXiv e-prints, p. 0904.2004

\bibitem[\protect\citeauthoryear{{Springel}, {Yoshida} \& {White}}{{Springel}
  et~al.}{2001}]{Springel2001}
{Springel} V.,  {Yoshida} N.,    {White} S.~D.~M.,  2001, New Astronomy, 6, 79

\bibitem[\protect\citeauthoryear{{Stamatellos}, {Whitworth}, {Bisbas} \&
  {Goodwin}}{{Stamatellos} et~al.}{2007}]{Stamatellos2007}
{Stamatellos} D.,  {Whitworth} A.~P.,  {Bisbas} T.,    {Goodwin} S.,  2007,
  \aap, 475, 37

\bibitem[\protect\citeauthoryear{{Testi} \& {Sargent}}{{Testi} \&
  {Sargent}}{1998}]{Testi1998}
{Testi} L.,  {Sargent} A.~I.,  1998, \apjl, 508, L91

\bibitem[\protect\citeauthoryear{{Vazquez-Semadeni}, {Ostriker}, {Passot},
  {Gammie} \& {Stone}}{{Vazquez-Semadeni} et~al.}{2000}]{Vazquez2000}
{Vazquez-Semadeni} E.,  {Ostriker} E.~C.,  {Passot} T.,  {Gammie} C.~F.,
  {Stone} J.~M.,  2000, Protostars and Planets IV, p.~3

\bibitem[\protect\citeauthoryear{{Walch}, {Burkert}, {Whitworth} A.and~{Naab}
  \& {Gritschneder}}{{Walch} et~al.}{2009}]{Walch2009}
{Walch} S.,  {Burkert} A.,  {Whitworth} A.and~{Naab} T.,    {Gritschneder} M.,
  2009, ArXiv e-prints, p. 0901.2127 {\bf (WO9) }

\bibitem[\protect\citeauthoryear{{Wetzstein}, {Nelson}, {Naab} \&
  {Burkert}}{{Wetzstein} et~al.}{2008}]{Wetzstein2008}
{Wetzstein} M.,  {Nelson} A.~F.,  {Naab} T.,    {Burkert} A.,  2008, ArXiv
  e-prints, 802, 0802.4245

\end{thebibliography}

\label{lastpage}
\end{document}